\documentclass[useAMS,usenatbib,usegraphicx,usenatbib]{mn2e}
\def\Zb{\overline{Z}}
\def\mb{\overline{m}}
\def\T1{{\rm 1T}}

\newcommand{\subs}[1]{{\rm _{s#1}}}
\newcommand{\sube}[1]{{\rm _{e#1}}}
\newcommand{\subi}[1]{{\rm _{i#1}}}
\newcommand{\subei}{_{\mathrm ei}}

\newcommand{\xs}{x\subs{}}
\newcommand{\vff}{v_{\mathrm{ff}}}

\newcommand{\rhoa}{\rho_{\rm a}}
\newcommand{\me}{m\sube{}}

\newcommand{\mproton}{m_{\mathrm{p}}}

\newcommand{\tG}{{\tilde \Gamma}}
\newcommand{\tL}{{\tilde \Lambda}}

\newcommand{\psic}{\psi_{\rm c}}
\newcommand{\psiei}{\psi_{\rm ei}}

\newcommand{\kb}{k_{_{\rm B}}}

\newcommand{\gb}{g_{_{\rm B}}}
\newcommand{\gB}{g_{_{\rm B}}}

\newcommand{\GS}{\ga}

\newcommand{\msun}{M_{_\odot}}

\title[Two-temperature accretion flows in mCVs]{
   Two-temperature accretion flows in magnetic cataclysmic variables: 
      Structures of post-shock emission regions and X-ray spectroscopy}
\author[C.~J.~Saxton et al.]
{ Curtis~J.~Saxton$^{1,2}$,
  Kinwah~Wu$^1$,
  Mark Cropper$^1$
  and Gavin Ramsay$^1$ \\
$^1$ Mullard Space Science Laboratory, University College London, 
   Holmbury St Mary, Dorking, Surrey RH5 6NT  \\
$^2$ Max-Planck-Institut f\"ur Radioastronomie, 
   Auf dem Huegel 68, D-53121 Bonn, Germany \\
}

\date{Received: }

\begin{document}
\twocolumn 

\maketitle

\begin{abstract}
We use a two-temperature hydrodynamical formulation  
  to determine the temperature and density structures 
  of the post-shock accretion flows in magnetic cataclysmic variables (mCVs) 
  and calculate the corresponding X-ray spectra.   
The effects of two-temperature flows are significant 
  for systems with a massive white dwarf and a strong white-dwarf magnetic field. 
Our calculations show that  
   two-temperature flows predict harder keV spectra than one-temperature flows 
   for the same white-dwarf mass and magnetic field. 
This result is insensitive to whether the electrons and ions 
   have equal temperature at the shock 
   but depends on the electron-ion exchange rate,  
   relative to the rate of radiative loss along the flow.  
White-dwarf masses obtained by fitting the X-ray spectra of mCVs 
   using hydrodynamic models including the two-temperature effects  
   will be lower than those obtained using single-temperature models.
The bias is more severe for systems with a massive white dwarf.     
\vspace*{0.25cm} 
\end{abstract} 


\section{Introduction}

Magnetic cataclysmic variables (mCVs) are close binaries 
  containing a magnetic white dwarf accreting material 
  from a Roche-lobe filling low-mass companion star
  \citep[see e.g.][]{warner1995,cropper1990}. 
The material flow is supersonic when it leaves the inner Lagrangian point 
  of the binary. 
It becomes subsonic near the white-dwarf surface,   
  and a shock is formed, 
  heating up the accreting material to temperatures  
  $T \approx 3GM_{\rm w}m_{_{\rm H}}/ 8kR_{\rm w} \sim 10 - 50$~keV 
  (where $G$ is the gravitational constant, $k$ the Boltzmann constant, 
  $m_{_{\rm H}}$ the hydrogen-atomic mass, 
  $M_{\rm w}$ the white-dwarf mass, 
  and $R_{\rm w}$ the white-dwarf radius).  
The accreting material in the pre-shock flow is thereby photoionised. 
The post-shock flow is cooled 
  by the emission of bremsstrahlung X-rays and cyclotron optical/IR radiation, 
  as the material settles onto the white-dwarf atmosphere.   
The X-ray emission from a mCV depends 
  on the temperature and density structures of the post-shock region, 
  which in turn depends on the properties, mainly the temperature, of the accretion shock.  
As the shock temperature is determined by
  the mass and radius of the accreting white dwarf, 
  we can infer the mass of the white dwarf from the X-ray spectra 
  \citep{rothschild1981,ishida1991,wu1995,fujimoto1997,ezuka1999}

The X-ray spectra of mCVs,
   in particular the subclass intermediate polars (IPs), 
   are well fitted by model spectra generated by
   the Aizu-type shock models \citep{aizu1973}, 
   such as those in \cite{chevalier1982,wu1994a,wu1994b}
   and \cite{cropper1999}.   
In these models, the electrons and ions have the same temperature locally
    (here termed a one-temperature model), 
    but the temperature and density change along the flow 
    in the post-shock region.   
It has been noticed that the white-dwarf masses
   obtained by the X-ray spectral fits
   using these models 
   tend to be systematically larger than those 
   derived from some other methods, e.g.\  optical spectroscopy 
   \citep[see][]{ramsay1998,ramsay2000}.

The discrepancies could be due to inaccuracies in these other determinations.
Alternatively, they may arise
   when the assumed absorption column density 
   in fitting the X-ray data is uncertain. 
They could also be due to 
   the fact that some relevant processes have not been considered 
   in deriving the temperature and density structures
   of the post-shock emission region.  
In the one-temperature model, 
   the electrons (the radiating particles) and the ions
   (the major energy-momentum carriers) 
   are strongly thermally coupled and share the same temperatures.  
The coupling is maintained by electron-ion collisions. 
However, if the radiative cooling timescale of the accretion flow 
   is shorter than the electron-ion collision timescale, 
   the electrons will lose their energy rapidly 
   while their energy gains via collisions
   are unable to keep up radiative loss.   
The two charged species then have unequal temperatures,   
  and the accretion flow becomes a two-temperature flow.      
\citep[See ][for recent reviews of accreting shocks in mCVs.]{wu2000,beuermann2004} 

Two-temperature flows are expected to occur 
   in mCVs containing a white dwarf with a very strong magnetic field
   ($B~\ga~30$~MG):
   \cite{lamb1979} 
   \citep[see also][]{imamura.thesis,imamura1987,imamura1996,
    woelk1996,saxton1999,saxton2001,fischer2001}. 
Cyclotron cooling is efficient in these systems,    
   and two-temperature effects are most significant 
   in the down-stream region just beneath  the accretion shock.    
The electron temperature and density structures
   of a two-temperature accretion flow 
   can be very different from those of a one-temperature flow.  
Moreover, the thicknesses of the high-density X-ray emitting region
   in the two flow models 
   are expected to differ as well. 
Thus, the mass estimates obtained 
  from a one-temperature post-shock model and
  a two-temperature post-shock model 
  could be different.  
  
Here we investigate the X-ray spectral properties of two-temperature  accretion flows in mCVs.  
We consider a semi-analytical approch, 
   in which the hydrodynamic models are constructed  
   following the prescriptions of  \cite{imamura1987} and \cite {saxton2001}.  
The thermal coupling between the ions and electrons 
  is parametrised by the Coulomb collision rate. 
The radiative loss is due to the emission of bremsstrahlung X-rays, 
   which is optically thin, and optical/IR cyclotron radiation, 
   which could have substantial optical depths.  
The total radiative loss is approximated 
   by a composite cooling function 
   as in the previous studies of one-temperature flows by  
  \cite{wu1994a,wu1994b,saxton.thesis}.
We calculate the temperature and density structures of 
   the two-temperature post-shock emission regions 
   and generate the X-ray line and continuum spectra (\S\ref{s.spectra}) 
  by convolving the MEKAL optically thin thermal plasmas model 
  \citep{mewe1985,kaastra1993} in the {\tt XSPEC} package.
The results of the two-temperature calculations 
  are compared with the results of one-temperature calculations 
  \citep[see e.g.][]{cropper1998,tennant1998,cropper1999}.

\section{Post-shock Accretion Flow: A Two-temperature Formulation}  

Our hydrodynamic formulation assumes that
    the gas in the post-shock region is completely ionised.
The flow is along the magnetic field lines. 
We omit the gravitational force and the curvature of the field lines. 
These effects could be important 
    when the thickness of post-shock region 
    is significant in comparison with the white-dwarf radius. 
    (See e.g.\ \citealp{cropper1999,canalle2005}.)
Thus, the flow is perpendicular to the white-dwarf surface 
   and is practically one-dimensional.  
Futhermore, we consider only the stationary situation.   
Hence, the hydrodynamic equations governing the flow are    
\begin{eqnarray}
  v {{\partial \rho} \over {\partial x}}\ 
       +\ \rho {{\partial v} \over {\partial  x}} & = & 0\ ,  \\
  {{\partial P_{\rm e}} \over {\partial x}}\ 
  +\  {{\partial P_{\rm i}} \over {\partial x}}\ 
       +\ \rho v {{\partial v} \over {\partial  x}} & = & 0\ , \\ 
   v {{\partial P_{\rm e}} \over {\partial x}}\
     -\ \gamma {{vP_{\rm e}} \over \rho} 
    {{\partial \rho} \over {\partial x}} 
        & = & (\gamma -1)~(\Gamma_{\rm ei} - \Lambda)\ , \\ 
   v {{\partial P_{\rm i}} \over {\partial x}}\ 
    -\ \gamma {{vP_{\rm i}} \over \rho} {{\partial \rho} \over {\partial x}} 
       & = & - (\gamma - 1)~\Gamma_{\rm ei}\      
\end{eqnarray}  
  where $v$ is the flow velocity, $\rho$ the density, 
  $P_{\rm e}$ the electron partial pressure, 
  $P_{\rm i}$ the ion partial pressure,  
  $\Gamma_{\rm ei}$ the rate of the electron-ion energy exchange, 
  $\Lambda$ the electron cooling function, 
  and $\gamma$ the adiabatic index. 
The total gas pressure $P$ 
  is the sum of the electron and ion partial pressures. 

We assume an ideal gas law for both the electron and ion gases, 
  i.e., $\gamma = 5/3$ and 
  $P_{\{\rm e,i\}} = n_{\{\rm e,i\}}~kT_{\{\rm e,i\}}$, 
  where $n_{\rm e}$ is the electron number density, 
  $n_{\rm i}$ the ion number density, 
  $T_{\rm e}$ the electron temperature, 
  and $T_{\rm i}$ the ion temperature.  
The rate of energy exchange due to electron-ion collision is 
  $\Gamma_{\rm ei} \approx 3 n_{\rm i} 
  k(T_{\rm i} - T_{\rm e})/ 2 t_{\rm ei}$, 
  where $t_{\rm ei}$ is the equi-partition time, given by 
\begin{equation}
  t_{\rm ei}\ =\  {{3 m_{\rm e} m_{\rm i} c^3} \over 
     {8~(2 \pi)^{1/2} Z_i^2 n_{\rm e} e^4~{\rm ln}~C}}~\bigl( 
    \theta_{\rm e} + \theta_{\rm i} \bigr)^{3/2}  
\end{equation} 
  \citep{spitzer1962}, where $m_{\rm e}$ is the electron mass, 
  $m_{\rm i}$ the ion mass,
  $c$ the speed of light, $e$ the electron charge, 
  $\theta_{\rm e} = k T_{\rm e} /m_{\rm e} c^2$, 
  $\theta_{\rm i} = k T_{\rm i} /m_{\rm i} c^2$, 
  and ln~$C$ the Coulomb logarithm. 
We have 
\begin{equation}
  \Gamma_{\rm ei} =  
   {{{4 \sqrt{2 \pi} e^4 Z_i^2 n_{\rm e} n_{\rm i}\ {\rm ln}~C}} 
   \over {m_{\rm e} c}}~\bigg\{ 
    {{ \theta_{\rm i} 
   [1- (m_{\rm e}\theta_{\rm e} / m_{\rm i} \theta_{\rm i})]} 
      \over {(\theta_{\rm e} + \theta_{\rm i})^{3/2}}}  \bigg\}
   \ .
\label{eq.Gamma}
\end{equation}  

The composite cooling function $\Lambda$  
   consists of a bremsstrahlung cooling term $\Lambda_{\rm br}$ 
   and a cyclotron cooling term $\Lambda_{\rm cy}$ 
   \citep{wu1994b}, i.e.,\ 
\begin{eqnarray} 
   \Lambda & \equiv &  \Lambda_{\rm br} +  \Lambda_{\rm cy} 
   \nonumber  \\  
      & \approx &  \Lambda_{\rm br}~\bigg[~1 + 
   \epsilon_{\rm s}~\bigg({T_{\rm e} \over T_{\rm e,s}} \bigg)^{2.0}  
      \bigg({n_{\rm e,s} \over n_{\rm e}} \bigg)^{1.85} \bigg]  \ ,    
\label{eq.cooling}
\end{eqnarray} 
  where $\epsilon \equiv t_{\rm br}/t_{\rm cy}$ 
  is the ratio of the bremsstrahlung-cooling timescale 
  to the cyclotron-cooling timescale. 
(Here and elsewhere, the subscript $s$ 
  denotes the value evaluated at the shock surface.)  
The derivation of the expression of $\epsilon_{\rm s}$ 
    is given in Appendix~\ref{appendix.cyclotron}. 
The explicit form of the bremsstrahlung cooling function is 
\begin{equation} 
\Lambda_{\rm br} = 16 \bigg({{2 \pi} \over 3} \bigg)^{3/2}
   { e^6  \over {m_{\rm e} c^2 h}}\ Z_i^2 n_{\rm e} n_{\rm i}\ 
   \theta_{\rm e}^{1/2}\ \gb
\label{eq.bremsstrahlung}
\end{equation} 
  \citep{rybicki},
  where $h$ is the Planck constant and 
  $g_{_{\rm B}}$ the Gaunt factor. 

We note that our calculations depend on the functional form of
   $\Lambda_{\rm cy}$
   in addition to the parameter $\epsilon_{\rm s}$.
In (\ref{eq.cooling}) we have assumed a power-law type function
   to approximate the cyclotron radiative loss term.
A key ingredient in constructing the cooling function is 
   to estimate the frequency $\omega^*$
   at which the local cyclotron spectrum peaks
   \citep[see][]{wada1980,saxton.thesis}.
How well the assumed power-law function approximates
   the cyclotron radiative loss depends on the accuracy of determining
$\omega^*$ in a given geometrical and hydrodynamic setting.
Here we follow the approach of \cite{wada1980} and \cite{saxton.thesis},
   but the technique would be improved
   if we could construct a local cyclotron cooling function
   more self-consistently,
   say by using an iterative scheme that calculates the
   cyclotron emission spectrum of each stratum
   and uses it to readjust
   the local parameters of the cyclotron cooling function.
  
We ignore electron conduction, Compton scattering 
   and nuclear burning in the energy transport.  
These processes are unimportant in the accretion flows of mCVs, 
  unless the situation is  very extreme, 
  e.g.\  the white dwarf is unusually massive ($\sim 1.2 - 1.4$~M$_\odot$) 
  and the accretion rate is very high ($\dot M > 0.1 \dot M_{_{\rm E}}$, 
  where $\dot M_{_{\rm E}}$ is the Eddington accretion rate) 
  \citep{imamura1987}. 
We do not include line cooling in our calculation of $\Lambda$. 
However, line cooling may not be negligible 
   at the very bottom of the post-shock region where the temperature is low. 
For systems with low white-dwarf masses,
   the shock temperature and the post-shock gas temperature are low  
   enough that the Fe~L lines can actually contribute a significant fraction
   of the total cooling  \citep[see][]{mukai2003}.
A fully consistent treatment of line cooling in the hydrodynamic calculation
   is non-trivial, and we will leave this for future studies. 

To simplify the hydrodynamic equations, 
  we consider the dimensionless variables 
  $\xi \equiv x/x_{\rm s}$, $\tau \equiv -v/v_{\rm ff}$, 
  $\zeta \equiv \rho/\rho_{\rm a}$, 
  $\pi_{\rm i} \equiv P_{\rm i}/\rho_{\rm a} v_{\rm ff}^2$, and 
  $\pi_{\rm e} \equiv P_{\rm e}/\rho_{\rm a} v_{\rm ff}^2$,   
  where $v_{_{\rm ff}} = (2 G M_{\rm w}/R_{\rm w})^{1/2}$ 
  is the free-fall velocity at the white-dwarf surface, 
  $\rho_{\rm a} = \dot m/ v_{\rm ff}$ 
  the density of the pre-shock flow, 
  and $\dot m$ the specific accretion rate. 
Substituting these variables 
  into equations (1), (2), (3) and (4) yields  
\begin{eqnarray}
  \tau {{\partial \zeta} \over {\partial \xi}}\ 
       +\ \zeta {{\partial \tau} \over {\partial \xi}} & = & 0,  \\
  {{\partial \pi_{\rm e}} \over {\partial \xi}}\ 
  +\  {{\partial \pi_{\rm i}} \over {\partial \xi}}\ 
       +\  \tau \zeta {{\partial \tau} \over {\partial \xi}} & = & 0, \\ 
   \tau {{\partial \pi_{\rm e}} \over {\partial \xi}}\
    -\ \gamma {{\tau \pi_{\rm e}}\over \zeta}{{\partial \zeta}
    \over {\partial \xi}} 
      & = & \tL - \tG_{\rm ei}\ , \\
   \tau {{\partial \pi_{\rm i}} \over {\partial \xi}}\ 
    -\ \gamma {{\tau \pi_{\rm i}}\over \zeta}{{\partial \zeta}
     \over {\partial \xi}} 
       & = &  {\tilde\Gamma}_{\rm ei}\ ,      
\end{eqnarray}   
where we define expressions for the non-dimensional 
  energy exchange and cooling functions,
  $\tG_{\rm ei}=(\gamma-1)(x_{\rm s}/\rho_{\rm a}v_{\rm ff}^3)~\Gamma_{\rm ei}$
  and 
  $\tL=(\gamma-1)(x_{\rm s}/\rho_{\rm a} v_{\rm ff}^3)~\Lambda$. 
 
The boundary values for electron and ion pressures 
  ($\pi_{\rm e,s}$ and $\pi_{\rm i,s}$)  
  are determined by the efficiency of the electron-ion coupling 
  through the shock transition region. 
Their ratio, $\sigma\subs{} \equiv \pi_{\rm e,s}/\pi_{\rm i,s}$, 
  can take values between $m_{\rm e}/m_{\rm i}$ 
  (the ratio of the electron mass to the ion mass) 
  and $\mu_{\rm i}/\mu_{\rm e}$
  (the ratio of the molecular weight of the ions 
  to that of the electrons), 
  depending on the assumed coupling processes 
  \citep{imamura1996}.
The physics of how the electrons couple with the ions 
   at the shock is not well understood.  
The value of $\sigma_{\rm s}$ is really
   a dependent property of the pre-shock flow,
   but its derivation would require solving a comprehensive
   model of pre- and post-shock regions,
   with explicit radiative transfer and hydrodynamics:
   a complex task beyond the scope of the present work.
We therefore treat  $\sigma\subs{}$ as a free parameter 
  and consider a few sensible values in our calculations 
  to see what difference it will make 
  to the density and temperature structures of the flows 
  and the associated X-ray spectral properties.   
  
For the other boundary conditions, 
  we assume a strong-shock condition: 
  at the shock surface ($\xi = 1$),  
  $\tau_{\rm s} = 1/4$, $\zeta_{\rm s} = 4$ and 
  for the (unitless) total gas pressure 
  ($\pi_0 \equiv \pi\sube{} + \pi\subi{}$) we have
  $\pi_{\rm 0,s} = 3/4$.
At the bottom of the flow, 
  we consider a ``stationary wall''  condition:  
  $\tau = 0$ at the white-dwarf surface ($\xi =0$). 
We omit the transfer of energy from beneath the white-dwarf atmosphere 
  \cite[see][]{wu2001}

\section{Temperature and Density Structures of the Post-shock Region}   

\begin{figure*}
\begin{center} 
\includegraphics[width=17cm]{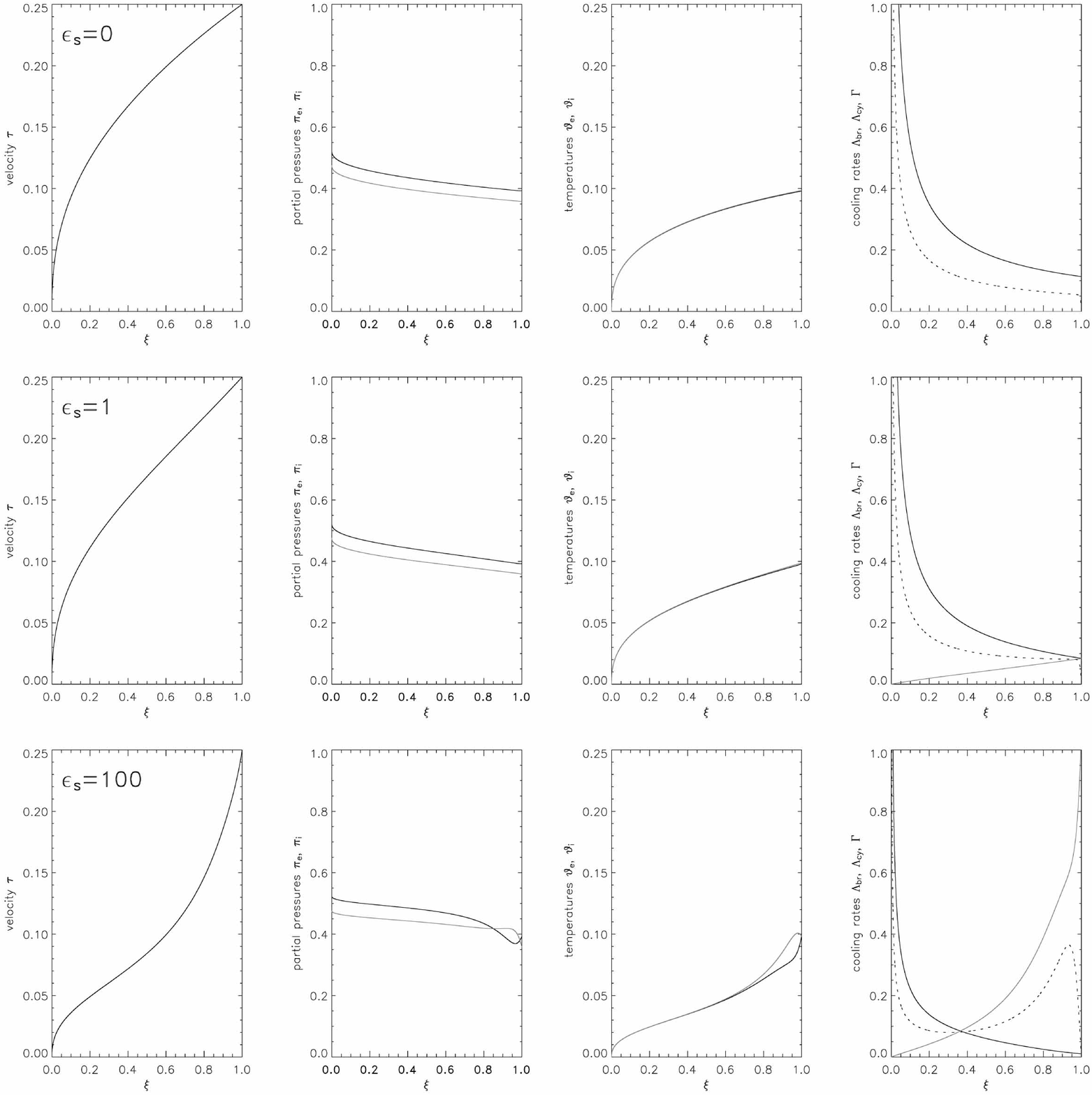}  
\end{center} 
\caption{ 
Stationary structures of accretion shocks 
  with $(\sigma\subs{},\psiei)=(\Zb,10)$
  and solar metallicity ($\Zb = {\overline {Z}}_\odot$).
The parameter that sets the relative efficiency 
  of cycloton cooling to bremsstrahlung cooling is
  $\epsilon\subs{} = 0,1,100$, from top to bottom. 
The first column shows the flow velocity $\tau$.
The second column shows the electron pressure $\pi\sube{}$ (black)
  and the ion pressure $\pi\subi{}$ (grey).
The third column shows 
  the electron temperature $\vartheta\sube{}\equiv\tau\pi\sube{}$ (black)
  and the ion temperature $\vartheta\subi{}\equiv\Zb\tau\pi\subi{}$ (grey).
The fourth column shows the local values 
  of the normalised bremsstrahlung cooling rate (black),
  the normalised cyclotron cooling rate (grey), 
  and the electron-ion exchange rate (dotted). 
These flows are quasi-one-temperature, 
   and the two-temperature case with small $\epsilon\subs{}$
   are practically indistinguishable from the one-temperature flows. }
\label{figure 1}
\end{figure*}

\begin{figure*}
\begin{center}
\includegraphics[width=17cm]{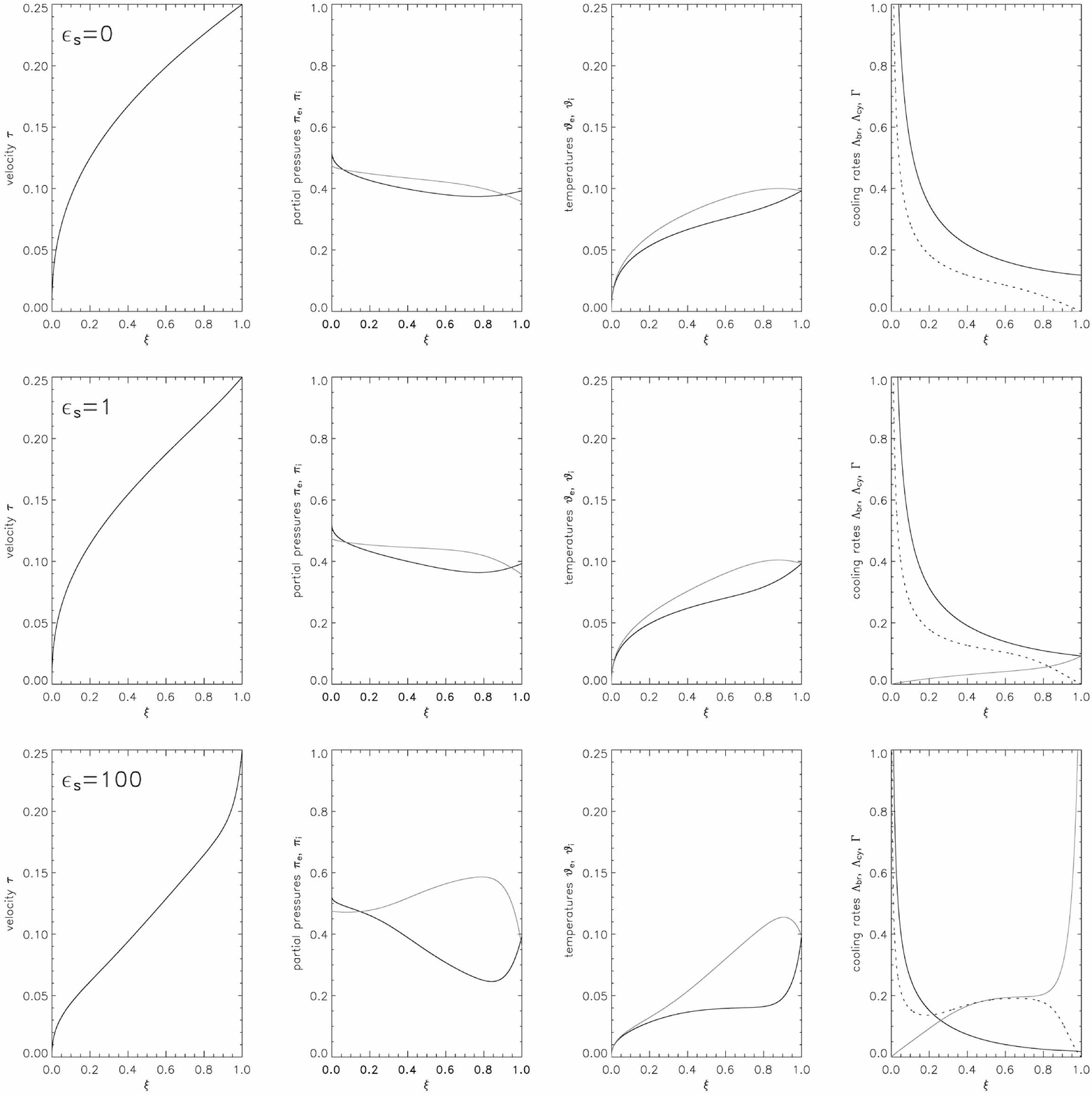}  
\end{center} 
\caption{
  Stationary structures of two-temperature accretion shock. 
The system parameters are the same as those in Figure~\ref{figure 1} except 
  $(\sigma\subs{},\psiei)=(\Zb, 0.2)$, 
  i.e. the electron-ion collisional coupling are weaker. } 
\label{figure 2}
\end{figure*}   

\begin{figure*}
\begin{center}
\includegraphics[width=17cm]{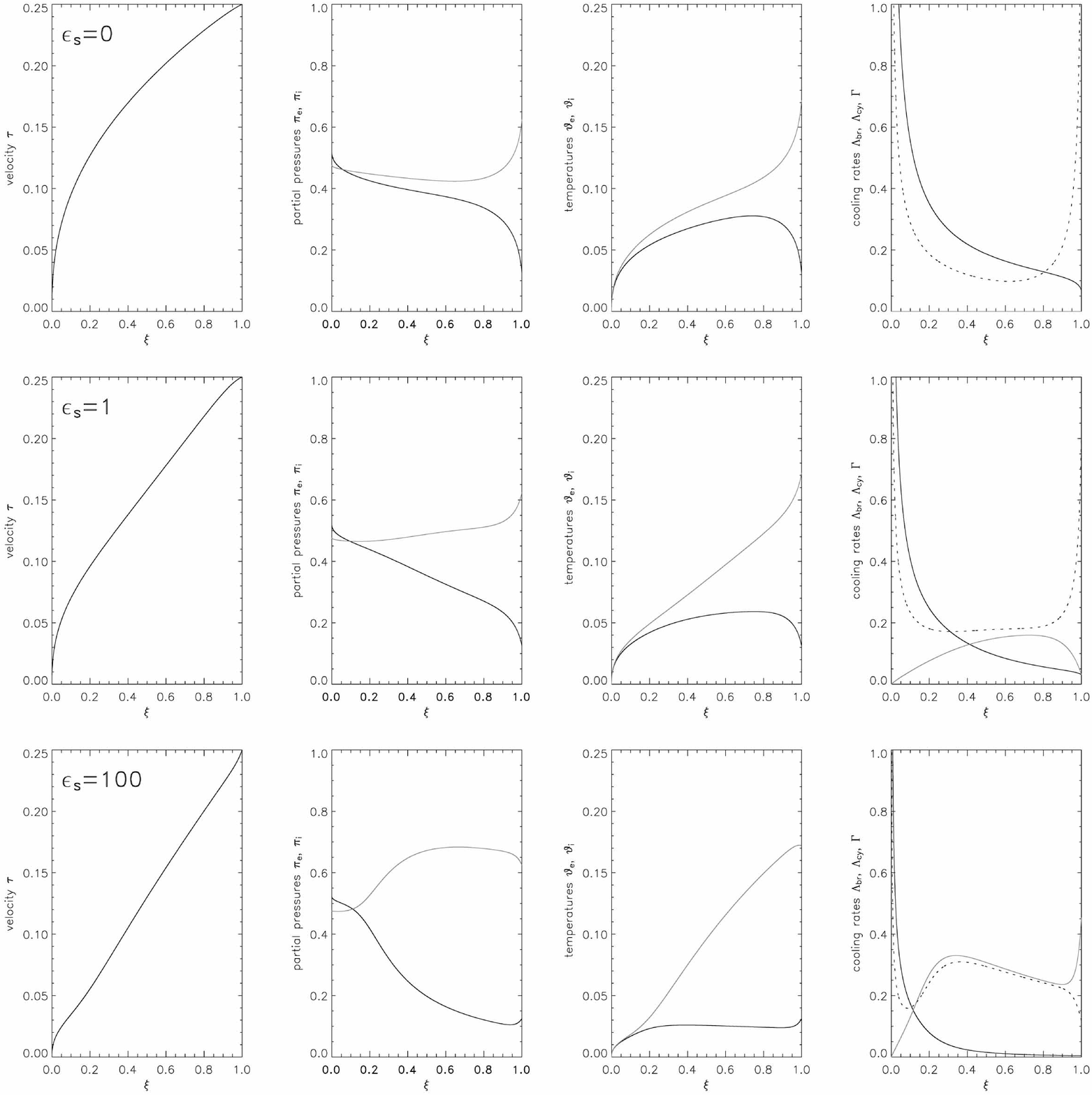}  
\end{center} 
\caption{
Stationary structures of two-temperature accretion shock. 
The system parameters are the same as those in 
  Figures~\ref{figure 1} and \ref{figure 2} except 
  $(\sigma\subs{},\psiei)=(0.2, 0.2)$.  }
\label{figure 3}
\end{figure*}   

\begin{figure*}
\begin{center}
\includegraphics[width=17cm]{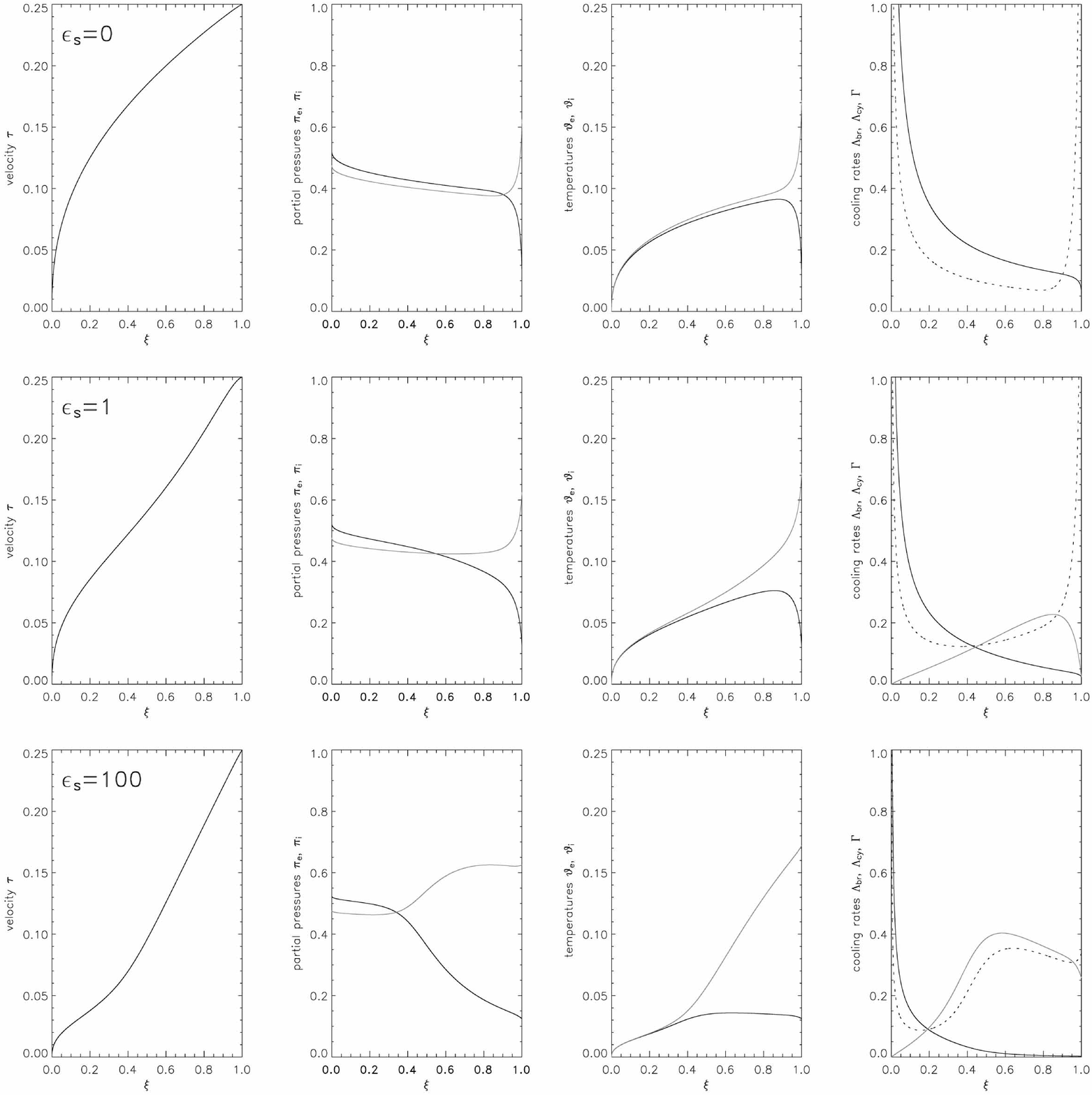}  
\end{center} 
\caption{
Stationary structures of two-temperature accretion shock. 
The system parameters are the same as those in 
  Figures~\ref{figure 1} and \ref{figure 2} except 
  $(\sigma\subs{},\psiei)=(0.2, 1)$.  }
\label{figure 4}
\end{figure*}

We integrate the mass continuity and the momentum equations, yielding   
\begin{eqnarray} 
  \tau  \zeta   & = &  1,  \\  
  \pi_{\rm 0}  & = &1 - \tau.  
\end{eqnarray}   
Substituting these into the energy equations 
  and eliminating $\pi_{\rm i}$, 
  we obtain two differential equations: 
\begin{equation} 
  {{\partial\tau} \over {\partial\xi}} =
     {{\tilde\Lambda}\over{ \gamma(1-\tau)-\tau }}
   \ ;
   \label{eq.diff.velocity}
\end{equation}
\begin{eqnarray}
  {{\partial\pi_{\rm e}} \over {\partial\xi}} =
  {1\over\tau} \left[{
      \tilde\Lambda
      -{{\gamma\tau\pi_{\rm e}\tilde\Lambda}\over{ \gamma(1-\tau)-\tau }}
      -\tilde\Gamma
   }\right]
   \ ,
   \label{eq.diff.electrons}
\end{eqnarray}  
  where 
\begin{equation}
  \tilde\Lambda = 
   (\gamma-1) x_{\rm s}\rho\vff^{-2} A
   \sqrt{ {\pi_{\rm e}}\over{\tau^3} }
    [1+\epsilon_{\rm s}f(\tau,\pi_{\rm e})]
   \ ,
\end{equation}
\begin{equation}
  \tilde\Gamma = (\gamma-1)x_{\rm s}\rhoa\vff^{-4} X
   {{1-\tau-\chi\pi_{\rm e}}\over{\tau^{5/2}\pi_{\rm e}^{3/2}}}
  \ ,
\end{equation}
the constants $A$ and $X$ depend on the composition of the plasma
(see Appendices~\ref{appendix.hydro}-\ref{appendix.psiei}),
\begin{equation}
  f(\tau,\pi\sube{})={{4^{\alpha+\beta}}\over{3^\alpha}}
  \left({ {1+\sigma\subs{}}\over{\sigma\subs{}} }\right)^{\alpha}
  \pi\sube{}^\alpha \tau^\beta
\end{equation}
  describes the efficiency of the secondary cooling process 
  (e.g.\ cyclotron cooling) relative to thermal bremsstrahlung cooling.
The constant $\chi = (\Zb+1)/\Zb$
  depends on the abundance-weighted mean charge of the ions.
We define a parameter $\psiei = X / A \vff^2$
  to express the efficiency of ion-electron energy exchange 
  compared to the radiative cooling.

The velocity, density, temperature and pressure profiles 
  of the ions and electrons in the post-shock flow 
  can be obtained by integrating
  equations (\ref{eq.diff.velocity}) and (\ref{eq.diff.electrons}). 

Assuming a neutral balance of electron and ion charges within the plasma,
  the one-temperature condition, $T_{\rm e}=T_{\rm i}$,
  implies a particular ratio of the partial pressures,
  $\pi_{\rm e}=\Zb\pi_{\rm i}$.
By setting $\gamma = 5/3$, $\sigma\subs{} = \Zb$ 
  (see equation \ref{eq.temp.ratio} in the appendices)
  and $\pi_{\rm e} = (1-\tau)/\chi$ in equation (\ref{eq.diff.velocity}), 
  we recover equation (2) in \cite{wu1994a}
  (with $\alpha = 2.0$ and $\beta = 3.85$)  
  for the one-temperature flows with
  a power-law approximation to 
  cyclotron cooling. 
If we assume the same expression 
  for $\Gamma$, $\sigma\subs{}$ and $\pi_{\rm e}$
  in equation (\ref{eq.diff.electrons}), 
  we will obtain an equation differing from equation (\ref{eq.diff.velocity}) 
  by a factor of $1/\chi$ in the terms at the left side. 
This is because the assumptions
   of charge neutrality and 
   equi-partition between the electron and ion energy
   (i.e.\ $\pi_{\rm e} = \Zb \pi_{\rm i}$) 
   in the entire shock-heated region requires the exchange term 
   to be determined by the electron cooling rate. 
An additional assumption of the energy-exchange rate depending on
   the difference between the ions' and the electrons' temperatures
   will inevitably lead to inconsistency. 
If we replace the energy-exchange term in equation (\ref{eq.diff.electrons}) 
  by $1/\chi$ of the cooling term, 
  then equations (\ref{eq.diff.velocity})
  and (\ref{eq.diff.electrons}) are identical 
  for $\gamma = 5/3$, $\sigma\subs{} = \Zb$ and $\pi_{\rm e} = (1-\tau)/\chi$.  

\subsection{Quasi-one-temperature flows}  

When the electron-ion coupling is strong (i.e.  the value of $\psiei$ is sufficiently large), 
   we expect the flow to be quasi-one-temperature. 
In Figure~\ref{figure 1} 
  we show the examples of flows which are effectively one-temperature 
  (with $\sigma\subs{} = \Zb$ and $\psiei =  10$) 
  and then how the flows deviate from the one-temperature limit 
  when we increase the cyclotron cooling rate.  
Cases shown are those of bremsstrahlung-only cooling 
  ($\epsilon\subs{}=0$),
  equally efficient bremsstrahlung and cyclotron cooling 
  at the shock ($\epsilon\subs{}=1$),
  and dominant cyclotron cooling 
  throughout much of the post-shock region ($\epsilon\subs{}=100$). 

For flows with small $\epsilon\subs{}$, 
  the electron and ion pressures have the same profiles 
  throughout the post-shock region.    
The temperatures of the electrons and ions 
   are indistinguishable, 
   and the flows are one-temperature. 
When $\epsilon\subs{}$ is large 
  (i.e. very efficient cyclotron cooling),  
  the electron and ion pressure have a different profiles 
  and  the electron temperature deviates from the ion temperature 
  in a small region near the shock.  
The flow velocity also deviates from one-temperature model near the shock.  
However, the temperatures of the electrons and ions eventually become the same 
  further downstream, 
  and the flow is effectively one-temperature in the base region
  of the two-temperature cases.    
Bremsstrahlung X-rays are emitted at the high-density regions at the bottom, 
   where the flow is practically one-temperature.  
We would therefore expect the X-ray properties to be similar 
   to those of the corresponding one-temperature cases.    
The optical/IR radiation from these flows 
   would be somewhat different
   from those of their one-temperature correspondences, 
   as cyclotron radiation
   (which occurs in the optical/IR wavelengths
   and has a substantial optical depth), 
   is emitted mainly from the hotter,
   less dense upper region of the post-shock flow.    

\subsection{Two-temperature flows} 

In general, 
  where electron-ion exchange is inefficient 
  compared to radiative cooling (i.e. small $\psiei$), 
  two-temperature effects become significant. 
Two-temperature flows can also occur 
  for moderately large $\psiei$, 
  if the initial difference between the electron and ion pressures 
  at the shock is substantial, i.e. small $\sigma\subs{}$.     
In Figures~\ref{figure 2}, \ref{figure 3} and \ref{figure 4} 
  we show three examples of  the two-temperature flows 
  with various combinations of the values for the parameters 
  $\sigma\subs{}$ and $\psiei$. 
   
For $(\sigma\subs{},\psiei)=(\Zb,0.2)$,    
  the collisional energy exchange with the ions
  does not keep pace with radiative loss of the electrons in the flow.    
Even though the electrons and ions 
  are set to have the equal temperatures at the shock,
  a strong disequilibrium ($T\sube{}<T\subi{}$)
  prevails throughout most of the post-shock region. 
Moreover, while the ions are shock-heated,
   the electron temperature does not rise accordingly 
   due to efficient radiative cooling and thermal decoupling with the ions.  
The difference between the electron and ion temperatures is obvious 
  for all three cases (with different $\epsilon\subs{} = 1$) that we consider, 
  and so is the difference between their electron pressure and ion pressure.  
When cyclotron cooling is weak ($\epsilon\subs{} \la 1$), 
  the velocity profiles of the two-temperature flow and one-temperature flows 
  are indistinguishable, 
  but  when cyclotron cooling is sufficiently strong
  (say $\epsilon\subs{} = 100$)
  the velocity profile deviates substantially from that of
  the one-temperature model. 
  
In all cases,  the density is high at the base of the post-shock region, 
   and collisional energy exchanges are more efficient than cyclotron loss.   
The collisional exchanges tend to bring the electrons and ions towards
   thermal equilibrium here. 
The bremsstrahlung X-rays are emitted copiously
   from the base of the post-shock flow. 
In a two-temperature flow, 
  the electrons and the ions gradually attain thermal equilibrium near the base 
   because the electron-ion collision rates increases with density, 
   and the flow become practically one-temperature again.  
In spite of this, 
  the electron temperature gradients near the white-dwarf surface 
  differ between the one-temperature and two-temperature flow models. 
These differences are, in some cases, sufficient to affect 
  the line and continuum X-ray spectra 
  (see \S\ref{s.spectra} and discussions in the later sections).

We note that the situation can be more complicated  
  in the regions near the shock. 
In the prescription that we consider, 
  the electron-ion exchange depends on the difference  
  between the electron and ion temperatures. 
Thus, the efficiency of the energy exchanges between the electrons and ions  
  is implicitly determined by the parameter $\sigma\subs{}$. 
Moreover,  the cyclotron cooling rate, 
  which depends mainly on the electron temperature 
  and is most effective in the hot region near the accretion shock, 
  is also limited by the efficiency of electron-ion exchange and hence $\sigma\subs{}$. 
The differences in properties of cyclotron emission   
  for the one-temperature and two-temperature flows should be noticeable.  
We can clearly see these effects in Figures~\ref{figure 3} and  \ref{figure 4} 
  (cf. Figures~\ref{figure 1} and \ref{figure 2}).   
The difference are more prominent 
  when the radiative cyclotron cooling rate at the shock, $\epsilon_{\rm s}$,
  is large.

\begin{table*}
\caption{ 
The total power of bremsstrahlung radiation
  $L_{\rm br}$, and cyclotron radiation $L_{\rm cy}$  
  (normalised to  $\rhoa\vff^3/x\subs{}$, 
  the kinetic energy of the pre-shock flow) 
  for different of $(\sigma\subs{},\psiei,\epsilon\subs{})$. 
In any case with $\epsilon\subs{}=0$,
  i.e. cyclotron cooling absent,
  we have $L_{\rm br}=0.5$ and $L_{\rm cy}=0$
  trivially.
} 
\hspace{15cm}
\begin{center}  
$ 
\begin{array}{ccc} 
\begin{array}{ccrcc}
\hline   
\sigma\subs{}&\psiei&\epsilon\subs{}&L_{\rm br}&L_{\rm cy}\\
\hline
0.2&0.1&  0.1&0.459&0.041\\
0.2&0.1&  1.0&0.346&0.154\\
0.2&0.1& 10.0&0.216&0.284\\
0.2&0.1&100.0&0.123&0.377\\
\\
0.2&0.2&  0.1&0.451&0.049\\
0.2&0.2&  1.0&0.330&0.170\\
0.2&0.2& 10.0&0.202&0.298\\
0.2&0.2&100.0&0.114&0.386\\
\\
0.2&0.5&  0.1&0.443&0.057\\
0.2&0.5&  1.0&0.316&0.184\\
0.2&0.5& 10.0&0.190&0.310\\
0.2&0.5&100.0&0.108&0.392\\
\\
0.2&1.0&  0.1&0.440&0.060\\
0.2&1.0&  1.0&0.309&0.191\\
0.2&1.0& 10.0&0.185&0.315\\
0.2&1.0&100.0&0.104&0.396\\
\hline
\end{array}  
& 
\begin{array}{ccrcc}
\hline   
\sigma\subs{}&\psiei&\epsilon\subs{}&L_{\rm br}&L_{\rm cy}\\
\hline
0.5&0.1&  0.1&0.486&0.014\\
0.5&0.1&  1.0&0.419&0.081\\
0.5&0.1& 10.0&0.288&0.212\\
0.5&0.1&100.0&0.171&0.329\\
\\
0.5&0.2&  0.1&0.484&0.016\\
0.5&0.2&  1.0&0.408&0.092\\
0.5&0.2& 10.0&0.273&0.227\\
0.5&0.2&100.0&0.161&0.339\\
\\
0.5&0.5&  0.1&0.481&0.019\\
0.5&0.5&  1.0&0.398&0.102\\
0.5&0.5& 10.0&0.260&0.240\\
0.5&0.5&100.0&0.152&0.348\\
\\
0.5&1.0&  0.1&0.480&0.020\\
0.5&1.0&  1.0&0.393&0.107\\
0.5&1.0& 10.0&0.254&0.246\\
0.5&1.0&100.0&0.147&0.353\\
\hline
\end{array}    
& 
\begin{array}{ccrcc}
\hline   
\sigma\subs{}&\psiei&\epsilon\subs{}&L_{\rm br}&L_{\rm cy}\\
\hline
1.0&0.1&  0.1&0.492&0.008\\
1.0&0.1&  1.0&0.446&0.054\\
1.0&0.1& 10.0&0.326&0.174\\
1.0&0.1&100.0&0.203&0.297\\
\\
1.0&0.2&  0.1&0.491&0.009\\
1.0&0.2&  1.0&0.441&0.059\\
1.0&0.2& 10.0&0.315&0.185\\
1.0&0.2&100.0&0.192&0.308\\
\\
1.0&0.5&  0.1&0.491&0.009\\
1.0&0.5&  1.0&0.436&0.064\\
1.0&0.5& 10.0&0.305&0.195\\
1.0&0.5&100.0&0.183&0.317\\
\\
1.0&1.0&  0.1&0.490&0.010\\
1.0&1.0&  1.0&0.434&0.066\\
1.0&1.0& 10.0&0.300&0.200\\
1.0&1.0&100.0&0.178&0.322\\
\hline
\end{array}
\end{array} 
$
\end{center}
\label{'table.parametric'}
\end{table*}

\subsection{Total radiative loss and X-ray luminosity}

In the hydrodynamic formulation that we consider
   and the boundary conditions that we adopt, 
   all the kinetic energy of the accreting gas 
   will be liberated via emitting bremsstrahlung X-rays
   and optical/IR cyclotron radiation. 
We now show that our prescription of the cooling function 
   ensures energy conservation 
   and hence self-consistency in the hydrodynamic calculations.   
   
The energy conservation requirement 
  that the total power radiated from the post-shock region
  equals the kinetic energy of the pre-shock flow,
  which is ${\frac12}\rhoa\vff^2$, 
  if no energy is transported across the white-dwarf surface.  
We can obtain the total power of radiation directly 
   by integrating the total cooling function 
   over the whole post-shock structure. 
In an explicit representation
  (with normalised density and velocity at the shock),  
\begin{eqnarray}
  L & = & \int_0^1   d\xi~ {{\tL}\over{\gamma-1}}
     \ =  \int_0^{1/4}  d\tau~ 
      \biggl[{{\gamma(1-\tau)-\tau }\over{\gamma-1}} \biggr] 
            \nonumber  \\ 
    & = & {{7\gamma-1}\over{32(\gamma-1)}} \ . 
\end{eqnarray}
Setting $\gamma=5/3$ 
  yields the the total power radiated via all processes $ L = 1/2$, 
  the result expected for exact energy conservation. 
    
The power of the bremsstrahlung X-rays  is 
\begin{equation}
L_{\rm br}
=\int_0^{1/4} d\tau~
  \biggl[ {{\gamma(1-\tau)-\tau }\over{\gamma-1}} \biggr]
 \biggl({1\over{1+\tL_{\rm cy}/\tL_{\rm br} }} \biggr) \ , 
\end{equation}
  and the power of cyclotron radiation is simply  
\begin{equation}
  L_{\rm cy} =L-L_{\rm br} \ . 
\end{equation}  
Values of $L_{\rm br}$ and $L_{\rm cy}$
  for different choices of the dimensionless system parameters 
  are shown in Table~\ref{'table.parametric'}.  
Values for representative choices of white dwarf mass,
  magnetic field and specific accretion rate are given 
  in Table~\ref{table.stellar}.

\begin{table*}
\caption{
Parameters of represenative accretion shock models,
  specified in terms of white dwarf mass, magnetic field strength
  and the election/ion pressure ratio at the shock, $\sigma_{\mathrm s}$.
We consider cases with $\sigma_{\rm s}=0.2$,
  $\sigma_{\rm s}=\Zb$ ($T_{\rm e}=T_{\rm i}$ at the shock),
  and a one-temperature model (1T) throughout the flow
  \citep{wu1994a}.
The following columns present
  corresponding values of the parameters $\psi_{\rm ei}$
  and $\epsilon_{\rm s}$,
  the bremsstrahlung and cyclotron luminosities 
  (in units of $\dot{m} \vff^2 / x_{\rm s}$),
  the shock height $x_{\rm s}$,
  and the electron temperature at the shock, $T_{\rm e,s}$.
The plasma composition is approximately solar
  and the mass flux is set to
  $\dot{m} = 1\ {\mathrm g}\ {\mathrm{cm}}^{-2}\ {\mathrm s}^{-1}$
  and the stream has cross-section $10^{15}\ {\mathrm{cm}}^2$.
}
\label{table.stellar}
\begin{center}
$
\begin{array}{ccccrrrcr}
\hline
 M/M_\odot & B/10\ {\mathrm{MG}}
 & \sigma_{\mathrm s} & \psi_{\mathrm{ei}} & \epsilon_{\mathrm s}
 & L_{\mathrm{br}} & L_{\mathrm{cy}} & x_{\mathrm s}/{\mathrm{cm}}
 & T_{\rm e,s}/{\rm keV}
\\
\hline
\\
0.5&  1& 0.2&     2.83&    0.0528	&0.461&0.039&2.08\times10^7&5.21\\
0.5&  3& 0.2&     2.83&    1.21    	&0.292&0.208&7.72\times10^6&5.21\\
0.5&  5& 0.2&     2.83&    5.18    	&0.210&0.290&3.74\times10^6&5.21\\
0.5&  1& \Zb&     2.83&    0.521  	&0.461&0.039&2.06\times10^7&16.4\\
0.5&  3& \Zb&     2.83&    11.9     	&0.291&0.209&7.48\times10^6&16.4\\
0.5&  5& \Zb&     2.83&    51.2     	&0.209&0.291&3.54\times10^6&16.4\\
0.5&  1& \T1&     \T1 &    0.521  	&0.460&0.040&2.05\times10^7&16.4\\
0.5&  3& \T1&     \T1 &    11.9     	&0.288&0.212&7.20\times10^6&16.4\\
0.5&  5& \T1&     \T1 &    51.2     	&0.206&0.293&3.28\times10^6&16.4\\
\\
0.7&  1& 0.2&     1.64&     0.262 	&0.389&0.111&3.32\times10^7&9.01\\
0.7&  3& 0.2&     1.64&     6.00   	&0.205&0.294&8.43\times10^6&9.01\\
0.7&  5& 0.2&     1.64&     25.7    	&0.144&0.356&3.99\times10^6&9.01\\
0.7&  1& \Zb&     1.64&     2.59   	&0.387&0.113&3.24\times10^7&28.3\\
0.7&  3& \Zb&     1.64&     59.2    	&0.204&0.296&7.75\times10^6&28.3\\
0.7&  5& \Zb&     1.64&     254      	&0.143&0.357&3.53\times10^6&28.3\\
0.7&  1& \T1&     \T1 &     2.47   	&0.384&0.115&3.15\times10^7&28.3\\
0.7&  3& \T1&     \T1 &     56.5    	&0.199&0.301&6.85\times10^6&28.3\\
0.7&  5& \T1&     \T1 &     242    	&0.139&0.361&2.87\times10^6&28.3\\
\\
1.0&  1& 0.2&     0.808&    2.06  	&0.267&0.233&4.50\times10^7&18.3\\
1.0&  3& 0.2&     0.808&    47.3  	&0.127&0.373&9.96\times10^6&18.3\\
1.0&  5& 0.2&     0.808&   203    	&0.088&0.412&4.84\times10^6&18.3\\
1.0&  1& \Zb&     0.808&   20.4   	&0.264&0.236&4.08\times10^7&57.4\\
1.0&  3& \Zb&     0.808&   467    	&0.126&0.374&8.16\times10^6&57.4\\
1.0&  5& \Zb&     0.808&   2002  	&0.087&0.413&3.77\times10^6&57.4\\
1.0&  1& \T1&     \T1  &   20.4   	&0.256&0.244&3.58\times10^7&57.4\\
1.0&  3& \T1&     \T1  &   467	         &0.120&0.380&5.67\times10^6&57.4\\
1.0&  5& \T1&     \T1  &   2002  	&0.083&0.417&2.26\times10^6&57.4\\
\\
\hline
\end{array}
$
\end{center}
\end{table*}

\begin{figure*}
\begin{center}
\includegraphics[width=17.5cm]{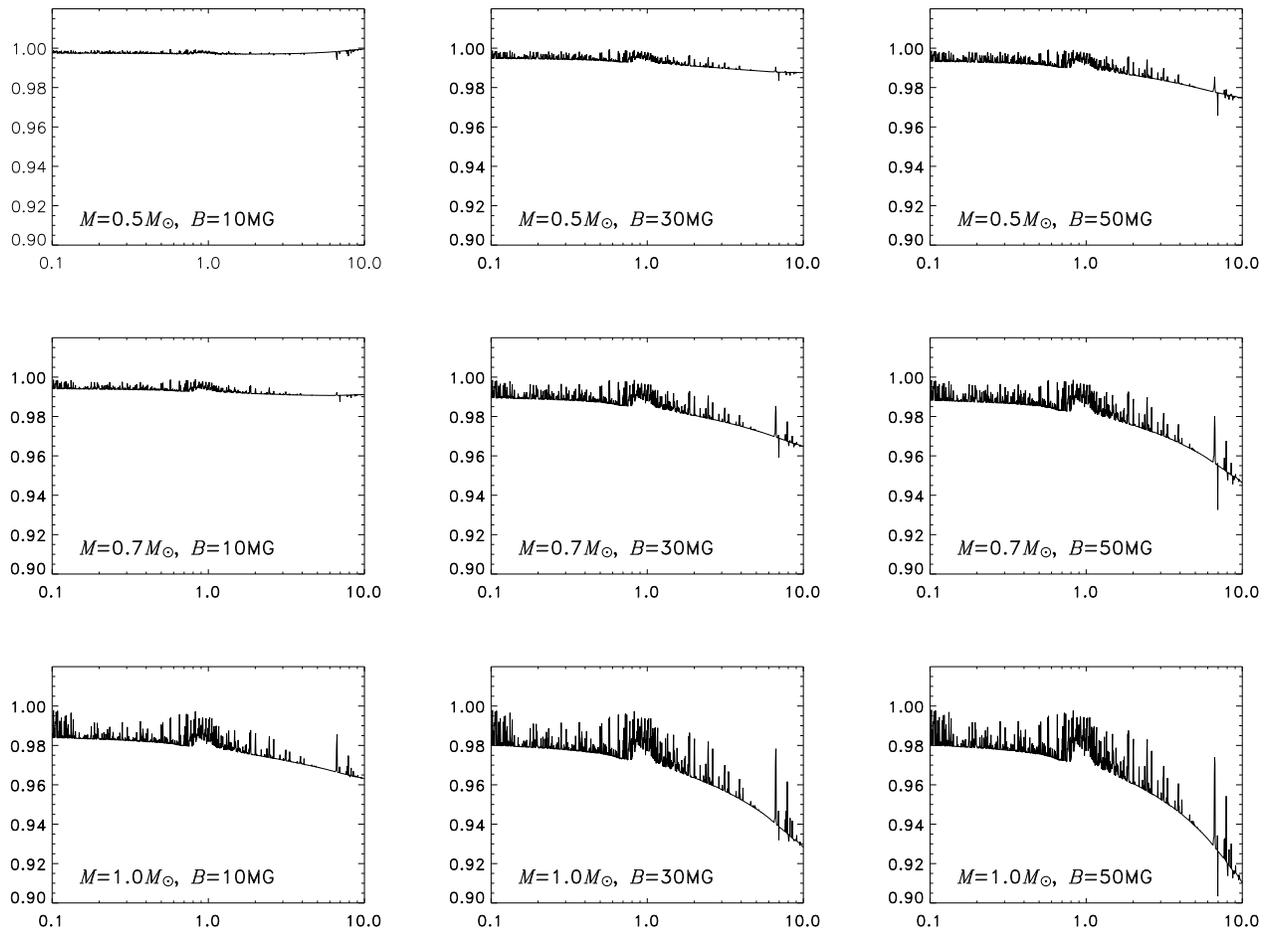}
\end{center}
\caption{
  The quotient X-ray spectra in the 0.1$-$10~keV band
      for various mCV parameters. 
  Each panel presents the ratio of
  the spectrum of the one-temperature flow model
      to the spectrum of the corresponding two-temperature flow model. 
  In all cases we set $\sigma_{\rm s}=\Zb$ and assume a solar metallicity.    
  The specific mass accretion rate is fixed to be
    $1.0~{\rm g}~{\rm cm}^{-2}{\rm s}^{-1}$.  
  The white-dwarf masses are  0.5, 0.7 and 1.0~$M_\odot$
    (from top row to bottom row). 
  The white-dwarf magnetic fields are 10, 30 and 50~$\mathrm{MG}$
    (from left column to right column). 
  In all cases, the two-temperature flow predicts
    a harder spectrum than a one-temperature flow. 
  The hardness is more severe when the white-dwarf mass
    and magnetic field increase.    
  }
\label{fig.xray.1}
\end{figure*}

\begin{figure*}
\begin{center}
\includegraphics[width=17.5cm]{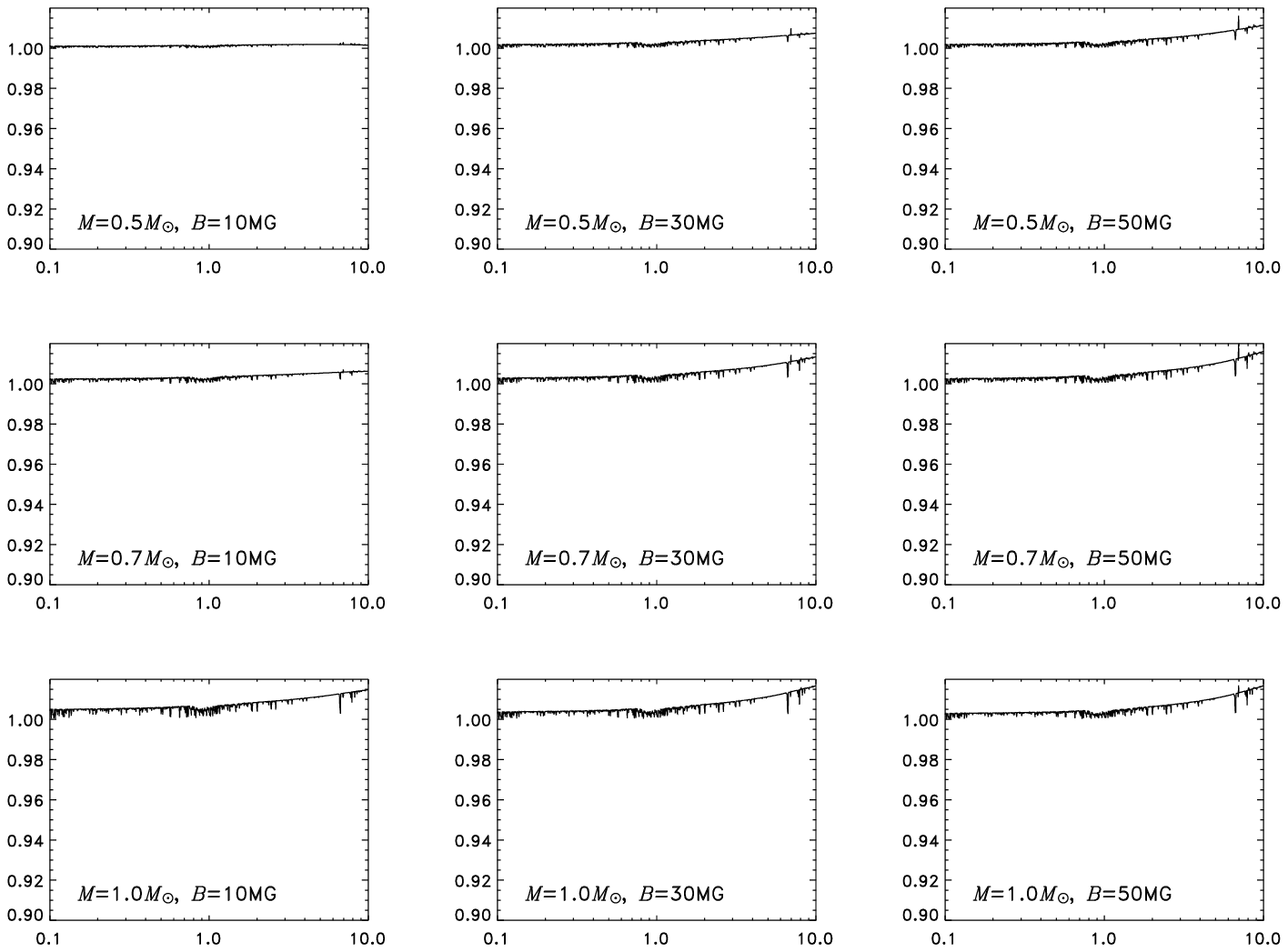}
\end{center}
\caption{
  The quotient X-ray spectra in the 0.1$-$10~keV band:
      the ratio of the two-temperature models with $\sigma_{\rm s}=0.2$
      to that of $\sigma_{\rm s}=\Zb$.
  The mCV parameters are equivalent to those in Figure~\ref{fig.xray.1}.
  Only the hard part of the spectrum is affected appreciably,
      with the case of $\sigma_{\rm s}=0.2$ yielding the harder results.
  The difference at 10~keV is at a level of less than $2\%$ for the case with
      greatest white dwarf mass and magnetic field intensity.
  }
\label{fig.xray.2}
\end{figure*}

\section{X-ray spectroscopy}
\label{s.spectra}

We now calculate the X-ray spectra of two-temperature flow models  
   and compare them with the spectra obtained 
   by the canonical one-temperature flow models.  
The question that we intend to answer is:
   how different are the spectral properties of two models
   for a given set of mCV parameters? 
More specifically, will the two-temperature flows produce harder X-ray spectra 
   than the one-temperature flows? 

We use the hydrodynamic formulation described  in the previous sections 
   to generate the density and temperature structures 
   of the post-shock emission regions,
   assuming appropriate mCV system parameters. 
In the spectral calculations 
   we adopt the same procedures as in \cite{cropper1999}.
We divide the post-shock emission region into a number of strata. 
The strata are each assumed to have constant density and electron temperature, 
   which take the corresponding mean values in the stratum.  
The radiative processes in the plasma are collisional-ionisation dominated 
    (we ignore photo-ionisation).  
Each stratum is assumed to be optically thin
   to the keV X-ray lines and continuum
   \citep[however see][]{wu2001b},
   and  we may use the MEKAL optically thin thermal plasma model    
   \citep{mewe1985,kaastra1993} in {\tt XSPEC}
   to calculate the local X-ray spectrum. 
The total spectrum is the sum of 
   the contributions from all the strata in the post-shock region
   (with the implicit, simplified assumption that line transfer and
   scattering effects are unimportant, cf. \citealt{kuncic2005}).
   
We consider white-dwarf masses of 0.5, 0.7 and 1.0 $M_\odot$, 
   and magnetic-field strengths of 10, 30 and 50~MG. 
The specific mass accretion rates $\dot m$ 
   is fixed to be 1.0~${\rm erg}~{\rm cm}^{-2}{\rm s}^{-1}$.  
The electron and ion temperatures are set to equal at the shock, 
  i.e. $\sigma_{\rm s}=\Zb$, 
  and a solar composition is used in determining $\Zb$ 
  and in generating the MEKAL spectra in {\tt XSPEC}.      
   
Given the complexity of the lines and the power-law like continuum, 
   the spectra of the two-temperature flows are not always easy
   to distinguish visually  from those of the one-temperature flows, 
We therefore consider the quotient spectra,  
   which are the ratios of the spactra of the one-temperature flows 
   to the spectra of the two-temperature flows.
   
Figure~\ref{fig.xray.1} shows the quotient spectra 
   (in the 0.1$-$10.0~keV band) for various combinations of system parameters. 
These spectra demonstrate that the competition between
  the radiative cooling of the electrons and the electron-ion energy exchange 
  affects the post-shock structure enough to influence the spectral properties, 
  and the effects are quite significant in some cases.
Generally, the one-temperature model predicts a softer X-ray spectrum
    for a given white dwarf mass.    
It also produces stronger emission lines, especially the Fe~L lines.   
These effects are more significant for greater white-dwarf masses. 

The softer X-ray emission of a one-temperature flow
    could be due to the fact that the one-temperature flow 
    tends to have a steep velocity gradient 
    in the mid-section of the post-shock region, 
    leading to a rapid increase in the electron and ion densities. 
As the X-ray emissivities are proportional to
    the square of matter density (for neutral plasmas), 
    this increases the relative contribution of the base region
    to the total  the emission. 
The more realistic two-temperature model 
   predicts a more gentle velocity gradient and hence density gradient, 
   and substantial X-ray lines and continuum
   are therefore emitted in the hotter strata of the post-shock region.  
When compared with a two-temperature flow,  
   the effective temperature of an one-temperature flow 
   is biased toward the cooler temperatures,  
   and as a consequence a softer spectrum results.  
The excess of the Fe~L lines in the one-temperature flow 
   can also be explained in the same manner. 
   
These differences imply that if we fit the observed spectrum of a mCV 
   with both models assuming the same magnetic field
   and specific mass accretion rate, 
   then the one-temperature model \citep{cropper1999}
   will give a higher white-dwarf mass than the two-temperature model.     
Thus, using the one-temperature model will over-estimate the white-dwarf mass, 
   and the bias is more severe for systems with
   a strongly magnetised, massive white dwarf.   
(We defer the estimation of observed systems' masses for future work;
   at some level of detail the results may be sensitive to
   the power-law approximation for cyclotron cooling, \S~2.)

Now an important question is: 
    how robust are the results given that we have assumed a specific
    $\sigma_{\rm s}$? 
While the value $\psiei$ is determined by $\Zb$ and $M_{\rm w}$
   (see Appendix~\ref{appendix.psiei}),
   we treat $\sigma\subs{}$ as a free parameter. 
We carry out calculations for two-temperature models with $\sigma_{\rm s}=0.2$ 
   and find that they produce spectra 
   which differ by less than 2\% from those 
   where $\sigma_{\rm s}=\Zb$ (see Figure~\ref{fig.xray.2}).

This result can be understood as follows:
although  $\sigma_{\rm s}$  can force the electrons and ions 
  to have unequal temperatures  at the shock,  
  it has weak effects on the flow in the base region.  
This can be seen by comparing the structure profiles 
  in the corresponding panels of Figures~\ref{figure 2} and \ref{figure 3},  
  which represent models that differ only in their values of $\sigma_{\rm s}$. 
The density and temperature profiles 
   approach the same values, with nearly the same spatial gradients, 
   near the white-dwarf surface.
Thus, the flow structure at the base 
   is insensitive to the electron-ion pressure ratio at the shock.
As the majority of the X-ray lines and continuum originate 
   from the bottom region near the white-dwarf surface,  
   the X-ray properties of the two cases should be very similar.  
Although we cannot use the observed X-ray spectrum 
   to infer the value of $\sigma_{\rm s}$ at the shock, 
   we can be certain that the mass estimate 
   depends on the effective shock temperature 
   and is less affected by the choice of $\sigma_{\rm s}$  
   in the spectral modelling.   
   
Inspection of Figure~\ref{fig.xray.1} reveals a `downward' spike 
   near the energy of Fe~K$\alpha$ lines, 
   which we identify as the emission from the H-like ions. 
This implies that the two-temperature flows and one-temperature flows
   predict very different ratios of the H-like (6.97~keV), He-like (6.7~keV)
   and neutral Fe~K$\alpha$ (6.4~keV) lines. 
The H-like Fe~K$\alpha$ line
   is weaker for the two-temperature flows
   than for the corresponding one-temperature flows,
   in the systems with a strongly magnetised white dwarf.   
This effect is stronger for more massive white dwarfs. 
The emission of H-like Fe~K$\alpha$ line requires a high plasma temperature
   ($\sim 10$~keV).  
The H-like Fe~K$\alpha$ lines is expected to originate 
   from regions closer to the shock,  
   which have higher temperatures than the region 
   that produces the He-like and neutral Fe~K$\alpha$ lines
   \citep[see e.g.][]{wu2001b}.
Thus, the H-like line does not share
   the characteristics of the lower ionised Fe lines. 
Nor, for the same reasons, does it share the characteristics
  of lines of the lighter metals, such as Si, S, Ar and O in the keV spectrum.
Moreover, as the emission region of the Fe~K$\alpha$ line
  is relatively near to the shock
  (where the difference between the electron and ion temperatures is greatest),
  the properties of the lines are sensitive to the assumed value of
  $\sigma_{\rm s}$ in the model (see Figure~\ref{fig.xray.2}).

\section{Conclusions}

We have presented a two-temperature hydrodynamical formulation  
  for accretion flows in mCVs 
  and used it to calculate the temperature and density structures 
  of the post-shock emission region. 
Two-temperature effects are significant 
   when the radiative loss is very efficient 
   such that electrons cannot acquire energy fast enough
   from the ions via collisions. 
We expect two-temperature flows to occur in systems 
   in which the white dwarf is massive and has a strong magnetic field. 
The magnetic field has less influence on the two-temperature effects
   if the white dwarf mass is small,
   but a relatively large effect for more massive white dwarfs.
(For example, contrast the flow structures in Figures~1 and 2;
   or in Figure~5 consider the greater sensitivity of the spectra
   in the bottom row where $M=1.0M_\odot$).
In high-mass, strong-field systems, cyclotron cooling is more efficient
   than bremsstrahlung cooling.  
In all cases, the flows eventually become one-temperature 
   near the base of the post-shock region, 
   where most of the X-rays are emitted. 

In spite of this, 
   two-temperature flows and one-temperature flows 
   have distinguishable X-ray properties 
   because of the differences in the density and temperature gradients     
   between the two flows.   
Our calculations show that  
   the X-ray spectra of one-temperature flows 
   are softer than the two-temperature flows, 
   if we assume the same system parameters, 
   such as white dwarf mass, magnetic field and specific mass accretion rate.  
This result is insensitive to the initial difference 
  between the electron and ion temperatures at the shock.  
White-dwarf masses of mCVs obtained 
   by fitting the X-ray spectra using a one-temperature flow model  
   will lead to overestimates, 
   especially for the systems with high white-dwarf masses and magnetic fields.

\section*{Acknowledgments} 

KW and CJS thank Klaus Beuermann for drawing our attention to
the significance of the functional form of the local cyclotron cooling.


\appendix

\section{System parameters}

\subsection{Hydrodynamic variables}
\label{appendix.hydro}

The gas is composed of electrons with mass $m\sube{}$
   plus species of ions with masses $m\subi{}$ and charges $Z_{\rm i}$.
If the ionic species have fractional abundances $f\subi{}$ by particle number
then the weighted mean ionic mass, charge and squared charge are
  $\mb\equiv\sum_{\rm i} f_{\rm i} m_{\rm i}$,
  $\Zb\equiv\sum_{\rm i} f_{\rm i} Z_{\rm i}$ and
  $\overline{Z^2}\equiv\sum_{\rm i} f_{\rm i}  Z_{\rm i}^2$.
For example,
  in the specific case of a completely ionised, purely hydrogen plasma
  these constants are $\mb=m_{\rm p}$, $\Zb=1$ and $\overline{Z^2}=1$.

In these terms, the number density of electrons is
\begin{equation}
n\sube =
\left(
{{\Zb}\over{\mb/\me +\Zb}}
\right)
{\rho\over\me}
\label{eq.n.e}
\end{equation}
and the number density of each ionic species is
\begin{equation}
n\subi =
\left(
{{f\subi{}}\over{\mb/\me +\Zb}}
\right)
{\rho\over\me}
= {{f\subi{}}\over{\Zb}} n\sube{}
\ ,
\label{eq.n.i}
\end{equation}
ensuring a balance of electric charge,
$n\sube{} = \sum_{\mathrm i}n\subi{} Z\subi{}$.

For a two-temperature shock, the electron and ion pressures are unequal.
The thermal variables, $\theta_{\{\mathrm{e,i}\}}$, are given by
\begin{equation}
\theta\sube{} \equiv {{k T\sube{} }\over{ \me c^2}}
= {{ P\sube{} }\over{ n\sube{} \me c^2 }}
= {{P\sube{}}\over{\rho c^2}}
\left({
{\mb/\me+\Zb}\over{\Zb}
}\right)
\ , \mbox{and}
\label{eq.theta.e}
\end{equation}
\begin{equation}
\theta\subi{} \equiv {{k T\subi{} }\over{ m\subi{} c^2}}
= {{ P\subi{} }\over{ n\subi{} m\subi{} c^2 }}
= {{P\subi{}}\over{\rho c^2}}
\left({
{\mb/\me+\Zb}\over{f\subi{} m\subi{}/\me}
}\right)
\ .
\label{eq.theta.i}
\end{equation}
At the shock surface we define a parameter
for the ratio of electron and ion partial pressures,
  $\sigma\subs{}\equiv P\sube{,s}/\sum_{\rm i} P\subi{,s}$,
  with summation over ion species ${\rm i}$.
For a strong shock, the total post-shock pressure equals
  $P=P\sube{}+\sum\subi{}P\subi{}={\frac34}\rhoa \vff^2$.
As the partial pressures are $P\sube{}=n\sube{}kT\sube{}$
   and $P\subi{}=n\subi{}kT\subi{}$,  
the temperatures of electrons and ions are related at the shock,
\begin{equation}
   n\sube{,s} T\sube{,s} = \sigma\subs{} \sum_{\rm i} n\subi{,s} T\subi{,s}.
\end{equation}
The post-shock electron temperature is
\begin{equation}
  T\sube{,s}={3\over{16}} {{m\sube{}\vff^2}\over\kb}
   \left({{\sigma\subs{}}\over{\sigma\subs{}+1} }\right)
   {{\Zb+\mb/m\sube{}}\over{\Zb}} \ . 
\label{'eq.2t.shock.temperature'}
\end{equation}
and the ion temperature (assumed to be the same for all i) is
\begin{equation}
   T\subi{,s}={{\Zb}\over{\sigma\subs{}}}T\sube{,s}
   \ .
\label{eq.temp.ratio}
\end{equation}

The upstream pre-shock velocity is assumed to be the
   free-fall velocity at the white-dwarf surface. 
We use the white dwarf mass-radius relation of \cite{nauenberg1972}.
The mass flux of the accretion column $\dot{m}$
   gives the pre-shock density $\rhoa=\dot{m}/\vff$.
The shock height $x\subs{}$ is calculated 
   by equating the bremsstrahlung cooling function at the shock
   with the realistic bremsstrahlung luminosity (involving $\rhoa$ and $\vff$),
   and substituting the numerically-determined normalisation constant $\psic$.

\subsection{Electron-ion exchange efficiency $\psiei$}
\label{appendix.psiei}

When (\ref{eq.n.e}), (\ref{eq.n.i}), (\ref{eq.theta.e}) and (\ref{eq.theta.i})
are substituted into (\ref{eq.Gamma})
and the $\Gamma_{\rm ei}$ functions are summed over ion species i
then it can be shown that
the total electron-ion energy exchange is
\begin{equation}
\Gamma = X \rho^{5/2} P\sube{}^{-3/2} (P - \chi P\sube{})
\end{equation}
where we define $\chi \equiv (\Zb+1)/\Zb$
and
\begin{eqnarray}
X&=&{{4\sqrt{2\pi}e^4}\over{\me^3 }} \ln C\
\left({
{\Zb}\over{\mb/\me+\Zb}
}\right)^{5/2}
\overline{ \left({ {\me Z^2}/{m} }\right) }
\nonumber \\
&\approx&
9.65\times10^{18}\ 
{{\ln C}\over{15}}
\left({
{\Zb}\over{ \mb/\me + \Zb }
}\right)^{5/2}
\overline{ Z^2/m }
\label{eq.X.cgs}
\end{eqnarray}
in c.g.s. units.
Equivalent substitutions in (\ref{eq.bremsstrahlung})
and summation over ion species i
leads to the total bremsstrahlung cooling function
\begin{equation}
\Lambda_{\mathrm br} = A \rho^2 (P\sube{}/\rho)^{1/2}
\end{equation}
where 
\begin{eqnarray}
A&= &
16 \left({ {2\pi}\over{3} }\right)^{3/2}
{{e^6}\over{\me^3 c^3 h}} \gB
{{\overline{Z^2}}\over{\Zb}}
\left({
{\Zb}\over{\mb/\me+\Zb}
}\right)^{3/2}
\nonumber \\
&\approx& 
5.61\times10^{16}
\gB \overline{Z^2}\ {\Zb}^{1/2}
\left({
{\mb+\Zb\me}\over\mproton
}\right)^{-3/2}
\label{eq.A.cgs}
\end{eqnarray}
in c.g.s units.

In a pure hydrogen plasma with equal electron and ion partial pressures
($P\sube{} = P/2$),
the bremsstrahlung cooling rate is
$\Lambda_{\mathrm br} \approx 3.97\times10^{16}
{\mathrm{erg}}\ {\mathrm{cm}}^{-3}\ {\mathrm s}^{-1}
(\rho / {\mathrm g}\ {\mathrm{cm}}^{-3})^{3/2}
( P / {\mathrm{dyn}}\ {\mathrm{cm}}^{-2} )^{1/2}$.
For solar plasma composition (i.e. $\Zb= \overline{Z}_\odot$), 
 we have 
$\Zb -1 = 0.09987$, 
$\overline{Z^2}=1.391$,
$\overline{Z^2/m}=6.007 \times 10^{23}\ {\rm g}^{-1}$,
$\overline{m}/\me=2366$. 
It follows that 
$A = 6.99\times10^{16}$
and $X=2.70 \times10^{34}$ in c.g.s units.
Thus the general effect of increasing metallicity
is to increase the efficiency of bremsstrahlung cooling,
which reduces the shock height if all else is equal.

The unitless form of the energy exchange function is
\begin{equation}
\tilde{\Gamma}
\equiv (\gamma-1) {{\xs}\over{\rhoa\vff^3}} \Gamma
= (\gamma-1) \psi_c \psi\subei
{{(1-\tau-\chi\pi\sube{})}\over{\tau^{5/2} \pi\sube{}^{3/2} }}
\ ,
\label{eq.Gamma.unitless}
\end{equation}
and the equivalent unitless function for the bremsstrahlung cooling is
\begin{equation}
\tilde{\Lambda}_{\mathrm br}
\equiv (\gamma-1) {{\xs}\over{\rhoa\vff^3}} \Lambda_{\mathrm br}
= (\gamma-1) \psi_c \tau^{-3/2} \pi\sube{}^{1/2}
\ .
\label{eq.Lambda.unitless}
\end{equation}
The dimensionless parameters $\psic$ and $\psiei$
\citep[defined in][]{imamura1996}
are constants of each accreting white dwarf system
in its particular accretion state.
Their values are,
in terms of the characteristics of the accretion flow
and universal physical constants,
given by:
\begin{equation}
\psiei \psic = X \xs \rhoa \vff^{-4}
\end{equation}
and
\begin{equation}
\psic = A \xs \rhoa \vff^{-2}
\ ,
\end{equation}
implying that
\begin{equation}
\psiei = {X\over{A \vff^2}}
\ ,
\end{equation}
which is purely a function of the white dwarf mass
   and composition of the accreting gas.
The model parameter $\psiei$ roughly describes
   the rate of electron-ion energy exchange
   compared to the rate of radiative cooling.
Upon substitution of (\ref{eq.X.cgs}) and (\ref{eq.A.cgs})
we have
\begin{equation}
\psiei = {{71.2}\over{\gB {v_8}^2}}
\left({ {\ln C}\over{15} }\right)
{{(\Zb)^2}\over{\overline{Z^2}}}
{{\mproton}\over{\mb+\Zb\me}}
\overline{ Z^2 \mproton / m }
\ ,
\label{eq.psiei.v8}
\end{equation}
where $v_8 = \vff / 10^8\ {\mathrm{cm}}\ {\mathrm s}^{-1}$.
For accreting magnetic white dwarfs $\ln C\approx15$ typically,
   and we take the Gaunt factor as $\gb\approx1.25$.
The factors that depend on gas composition 
   tend to lower the value of $\psiei$
   for accretion flows of higher metallicity.
Solar abundances imply a value of $\psiei$ roughly 32\% smaller
   than for pure hydrogen.

\begin{figure}
\begin{center}
\includegraphics[width=6cm]{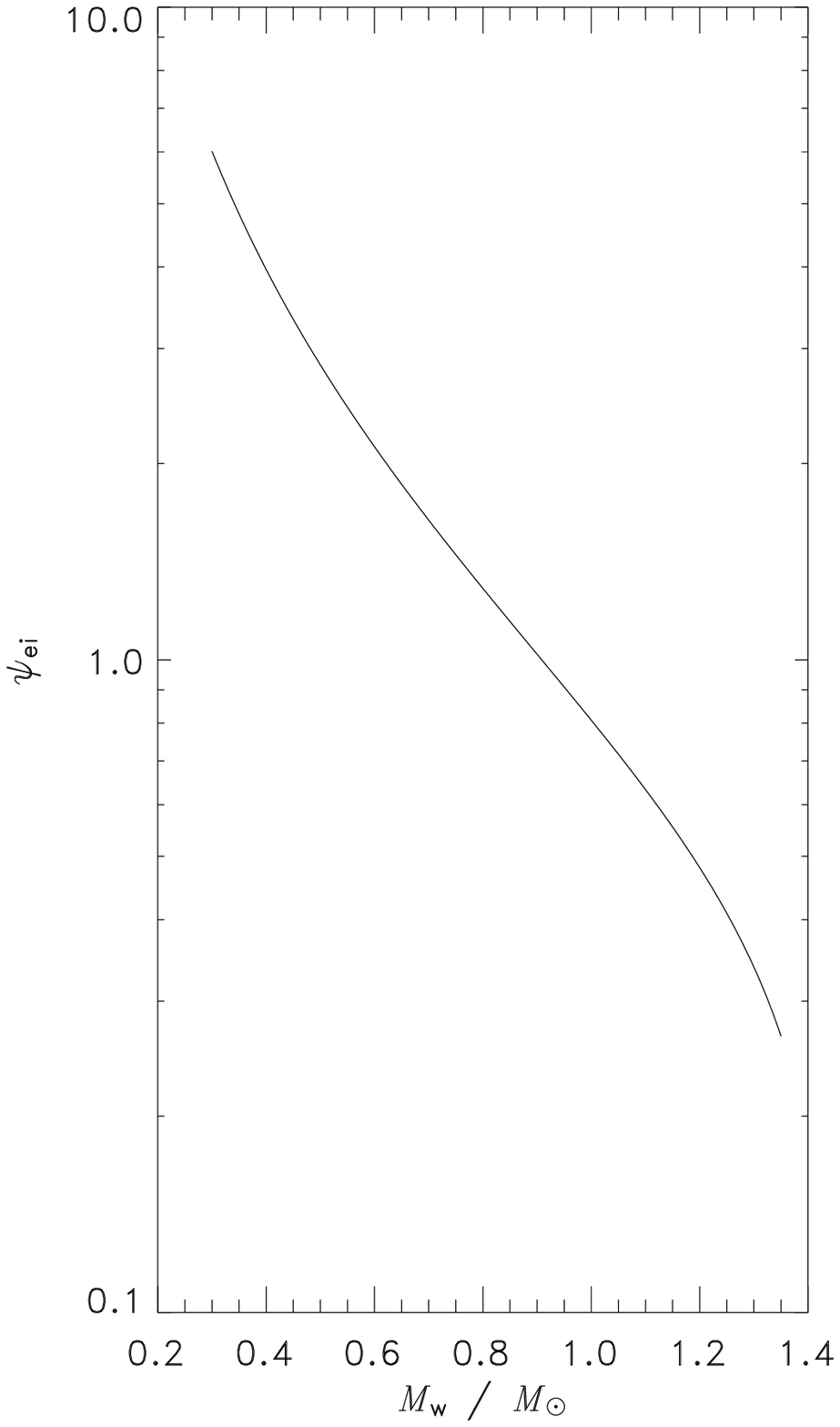}
\end{center}
\caption{
   Electron-ion exchange efficiency parameter $\psiei$
      verses white-dwarf mass.
   We assume approximately solar abundances.
   This parameter is independent of
      the magnetic field strength and $\sigma_{\rm s}$.
}
\label{fig.psiei}
\end{figure}

Using our assumption of solar abundances,
  the free-fall velocity must be greater than
  $\vff>6.21\times10^8~{\rm cm}~{\rm s}^{-1}$
  in order to make $\psiei<1$,
  i.e. radiative cooling comparable to, or more efficient than
  the exchange of thermal energy between electrons and ions.
For white dwarfs with masses $\GS0.9\msun$, the values of $\psiei$ are below 1,
  and become as low as $0.3$ for very massive white dwarfs
  with $\vff=1.13\times10^9~{\rm cm}~{\rm s}^{-1}$.
To obtain an extreme value of $\psiei=0.1$,
  a free-fall velocity of $\vff=1.96\times10^{9}~{\rm cm}~{\rm s}^{-1}$
  is required.
In this paper we consider values of orders $\psiei\sim0.1 - 10$,
  as calculated in Figure~\ref{fig.psiei},
  using the white dwarf mass-radius relation of 
  \cite{nauenberg1972}
  and mean molecular weight $\mu_{\rm w}=2.00$.
In \cite{imamura1996}, values of $\psiei$ 
  ranging from 0.01 to 1.0 were, however, used. 
We note that for white dwarfs in mCVs,
  values of $\psiei$ much greater than $0.01$ are necessary.

\subsection{Cyclotron cooling efficiency $\epsilon\subs{}$}
\label{appendix.cyclotron}

The timescales of the bremsstrahlung and cyclotron cooling processes
   are respectively 
\begin{equation}
  t_{\rm br}={\frac32} {{(n\sube{}+n\subi{})\kb T\sube{} }\over{\Lambda_{\rm br}}}
\end{equation}
\begin{equation}
  t_{\rm cy}={\frac32} {{(n\sube{}+n\subi{})\kb T\sube{} }\over{\Lambda_{\rm cy}}} \ .
\end{equation}
It is useful to compare the cooling timescales
   to express the local efficiency of cyclotron cooling 
   with respect to bremsstrahlung cooling.
The ratio of timescales at any given position in the post-shock flow
   will be written as 
\begin{equation}
  \epsilon(x)\equiv{{t_{\rm br}}\over{t_{\rm cy}}}
       ={{\Lambda_{\rm cy}}\over{\Lambda_{\rm br}}} \ .
\label{'eq.definition.epsilon'}
\end{equation}

This leads to the construction of $\epsilon\subs{}$,
  the relative efficiency of the cyclotron cooling
  as evaluated at the shock surface,
  which is an important dimensionless physical parameter
  used for the description and analysis of cooling accretion flows
  in which both bremsstrahlung and cyclotron processes are present.
Assuming that the cross-section of the flow is circular,
  $\epsilon\subs{}$ is obtained in realistic terms 
  by substituting appropriate system parameters and shock conditions
  into the equations for the $\Lambda$ cooling functions and
  (\ref{'eq.definition.epsilon'}).

In general,
   the efficiency of cyclotron cooling relative to bremsstrahlung cooling, 
   in terms of electron temperatures and number densities, is
\begin{equation}
\epsilon=0.0762 {{\Zb}\over{\gb\overline{Z^2} }}
   a_{15}^{-0.425} B_7^{2.85} n\sube{16}^{-1.85} T_8^2 \ .
\label{eq.epsilon.general}
\end{equation}
Since a circle is the two-dimensional geometric shape  
  with the minimum ratio of perimeter to internal area,
  the above expression for $\epsilon$ is actually a lower limit.
For more realistic cross-sections,
  the numerical factor $0.0762$ would become larger. 

\begin{figure}
\begin{center}
\includegraphics[width=6cm]{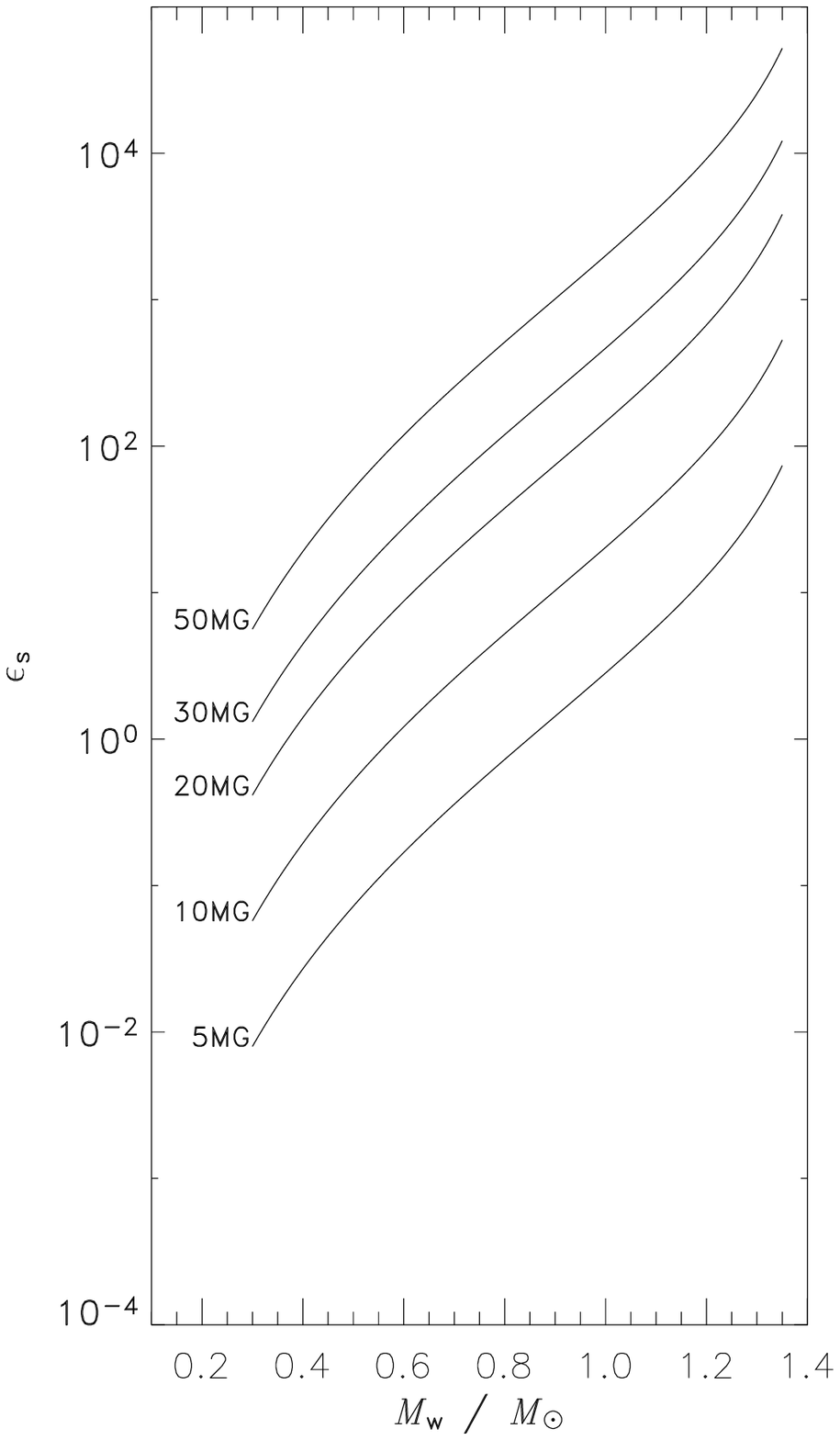}
\end{center}
\caption{
Cyclotron efficiency parameter $\epsilon\subs{}$
as a function of white-dwarf mass,
given for five choices of magnetic field strength.
Here we set the ratio of pressures at the shock as
$\sigma\subs{}=\overline{Z}$;
the mass accretion rate is $\dot{m}=1~{\rm g}~{\rm cm}^{-2}~{\rm s}^{-1}$;
the pole area is $a=1.0\times10^{15}~{\rm cm}^2$.
The flow cross-section is assumed to be circular,
and the composition is solar.
}
\label{fig.epsilon.Z}
\end{figure}

\begin{figure}
\begin{center}
\includegraphics[width=6cm]{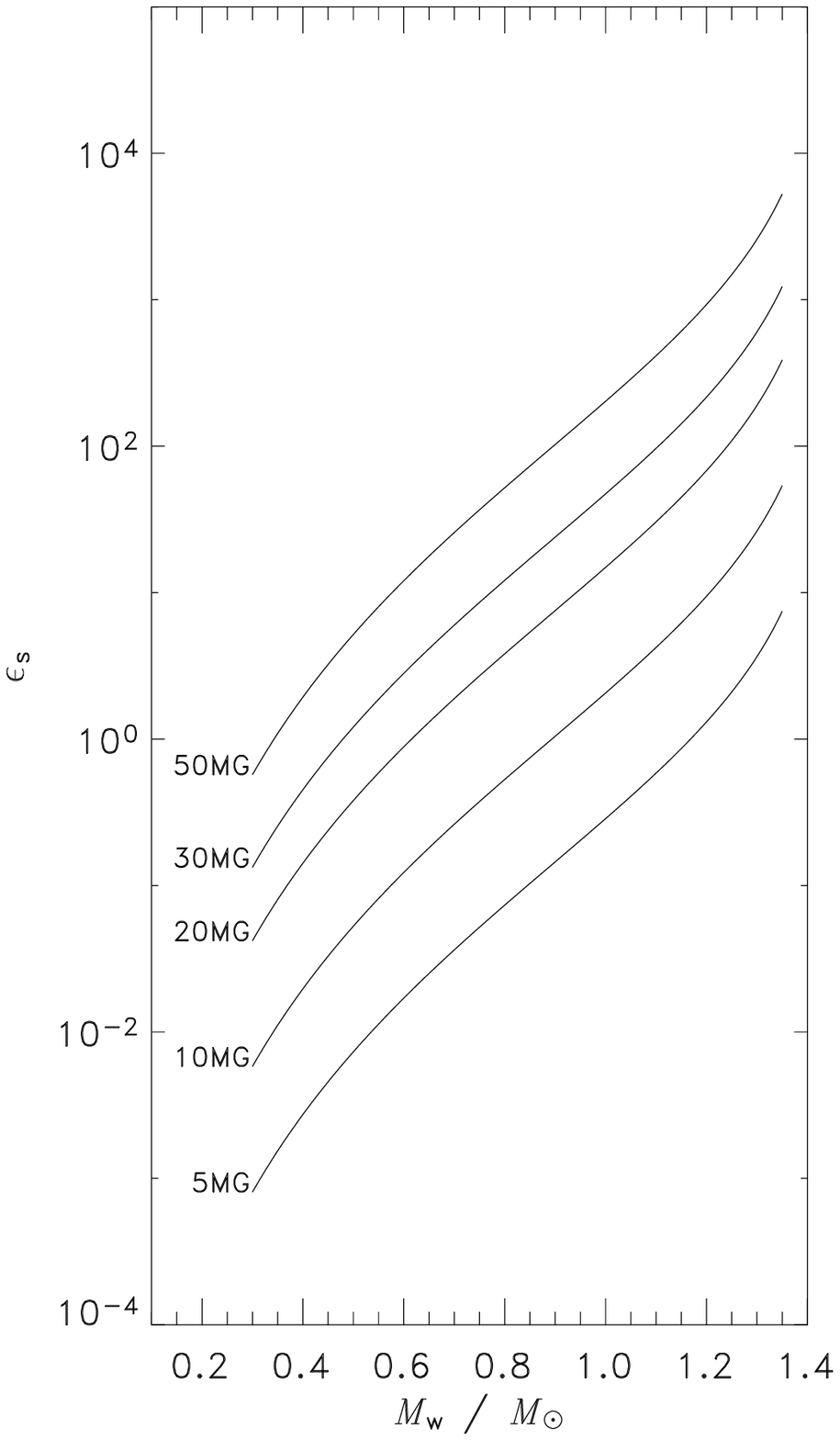}
\end{center}
\caption{
Same as Figure~\ref{fig.epsilon.Z} but with $\sigma\subs{}=0.2$.
}
\label{fig.epsilon.2}
\end{figure}

By substituting expressions 
  relating $n\sube{16}$ and temperature (\ref{'eq.2t.shock.temperature'})
  to the pre-shock density $\rhoa$ and free-fall velocity $\vff$,
  the efficiency $\epsilon\subs{}$ can be re-expressed  
  in terms of the properties of the white-dwarf accretion flow.
In the general case with multiple species of ions
\begin{eqnarray}
  \epsilon\subs{} & = & 
  {{2.13\times10^{-16}(\Zb+\mb/m\sube{})^{3.85}}
  \over{\gb\Zb^{2.85}\overline{Z^2}\left({1+\sigma\subs{}^{-1} }\right)^2 }} 
  \nonumber \\ 
  &  & \hspace*{1.5cm} 
  \times a_{15}^{-0.425} B_7^{2.85} \rho_{\rm a-8}^{-1.85} v_8^4 \ , 
  \label{eq.explicit.epsilon_s}
\end{eqnarray}
    where $\rho_{\rm a-8}\equiv\rhoa/{10^{-8}{\rm g}\cdot{\rm cm}^{-3}}$
    and $v_8\equiv\vff/{10^{8}{\rm cm}\cdot{\rm s}^{-1}}$.
This expression is slightly different from 
   the cyclotron/bremsstrahlung efficiency ratio 
   given in \cite{langer1982}
   where the emission region is semi-infinite but without a specific geometry.
Assuming a circular flow cross-section and parameters 
   appropriate for accretion shocks in AM~Herculis systems,
   with $a_{15}=1$ and $\dot{m}=1~{\rm g}~{\rm cm}^{-2}{\rm s}^{-1}$,
   the cyclotron-cooling efficiency parameter varies with
   white-dwarf mass and magnetic field as shown in
Figures~\ref{fig.epsilon.Z} and \ref{fig.epsilon.2}.

\subsection{System geometry}
\label{appendix.geometry}

The geometry of the accretion stream could be important 
  as optically thick cyclotron cooling depends on the emitting surface area.
Under our approximation, 
  (\ref{eq.epsilon.general}),
  the efficiency parameter $\epsilon\subs{}$ 
  is proportional to the ratio of perimeter to cross-sectional area.

A circular cross section for the accretion column 
  gives the minimum value of $\epsilon\subs{}$.
All else being equal,
  flows with cross-sections departing from a circle
  cannot have lower values of $\epsilon\subs{}$
  than given by (\ref{eq.explicit.epsilon_s}).
The greater the non-circularity, the greater $\epsilon\subs{}$ must be.
There is observational evidence 
  that in AM~Herculis type systems (polars)
  the cross-section of the accretion column near the the white-dwarf surface  
  is a banana-like arc
  \citep{cropper1985,wickramasinghe1988,potter1998}.
This geometry ought to give a $\epsilon\subs{}$ value several times higher.
Whether the real banana-shaped cross-sections have more crenulated or convoluted edges
  at finer spatial scales is unknown.
If they do, then the equivalent $\epsilon\subs{}$ values
  may be considerably greater than 
  the lower limit of the circular approximation.
Because of the effect of the surface area to volume ratio in a system radiating 
  via an optically thick process,
  real effective values of $\epsilon\subs{}$
  may be less sensitive to the magnitude of the polar magnetic field
  than to other influences determining the flow structure, 
  e.g. the geometric relationship between
  the magnetosphere and the companion star,
  and the condition of the flow where it threads onto the magnetic field.
If flow geometries are sufficiently diverse among mCV systems
  then there may exist some strong-field systems 
  which have lower $\epsilon\subs{}$ than
  weaker-field systems that have more circular flow sections.
The magnetic field strength is measurable,
  implying a lower limit on $\epsilon\subs{}$ in each system,
  but unfortunately the value of $\epsilon\subs{}$
  is not a directly observable quantity.

\end{document}